\documentclass{aa}  
\usepackage{CJK}
\usepackage[colorlinks=true,linkcolor=blue,citecolor=blue, urlcolor=blue]{hyperref}
\usepackage{natbib}
\usepackage{graphicx}
\usepackage{txfonts}
\usepackage{xcolor}
\usepackage[switch]{lineno}
\usepackage{orcidlink}

\def\FeH{{\rm[Fe/H]}}

\def\AA{$\; \buildrel \circ \over {\rm A}$}

\def\deg{^\circ}
\def\s{\ifmmode \widetilde \else \~\fi}
\def\={\overline}

\def\spose#1{\hbox to 0pt{#1\hss}}

\def\eg{{e.g.,\ }}
\def\ie{{i.e.,\ }}
\def\lta{\mathrel{\spose{\lower 3pt\hbox{$\mathchar"218$}}
     \raise 2.0pt\hbox{$\mathchar"13C$}}}
\def\gta{\mathrel{\spose{\lower 3pt\hbox{$\mathchar"218$}}
     \raise 2.0pt\hbox{$\mathchar"13E$}}}
\def\Dt{\spose{\raise 1.5ex\hbox{\hskip3pt$\mathchar"201$}}}    % upper case
\def\dt{\spose{\raise 1.0ex\hbox{\hskip2pt$\mathchar"201$}}}    % lower case

\def\ie{{ i.e.,\ }}
\def\eg{{ e.g.,\ }}

\newcommand{\gkai}[1]{\begin{CJK*}{UTF8}{gkai}\raisebox{.1em}{(}#1\raisebox{.1em}{)}\end{CJK*}}
\newcommand{\feh}{\ensuremath{\mathrm{[Fe/H]}}}

\newcommand{\Gaia}{\textit{Gaia}\xspace}
\newcommand{\CaHK}{\emph{CaHK}\xspace}
\defcitealias{Martin2024}{DR1}

\makeatletter

\begin{document}

   \title{The Pristine survey}
   \subtitle{XXIX. Data Release 2: Photometric metallicities for 15 millions stars down to $G\sim 20.5$}

    \titlerunning{Pristine DR2}
    \authorrunning{Yuan et al.}

   \author{Zhen~Yuan\thanks{Corresponding author: \email{zhen.yuan@nju.edu.cn}} \gkai{袁珍}\inst{1,2}\orcidlink{0000-0002-8129-5415}, Anke~Ardern-Arentsen\inst{3}\orcidlink{0000-0002-0544-2217}, Nicolas~F.~Martin \inst{4,5}\orcidlink{0000-0002-1349-202X}, Else~Starkenburg \inst{6}\orcidlink{0000-0003-4501-103X}, Morgan~Fouesneau \inst{5}\orcidlink{0000-0001-9256-5516}, Piercarlo~Bonifacio \inst{7}\orcidlink{0000-0002-1014-0635}, Elisabetta~Caffau \inst{7}\orcidlink{0000-0001-6011-6134}, Vanessa~Hill\inst{8}, Georges~Kordopatis \inst{8}\orcidlink{0000-0002-9035-3920}, Camila~Navarrete \inst{8,9,10}\orcidlink{0000-0002-4777-9934}, Julio~Navarro \inst{11,12}\orcidlink{0000-0003-3862-5076}, 
   Samuel Rusterucci\inst{4,6},
   Federico~Sestito \inst{13}\orcidlink{0000-0002-3182-3574}, Salvatore~Taibi\inst{14}\orcidlink{0000-0001-6469-8805}, Kim~A.~Venn\inst{11}\orcidlink{0000-0003-4134-2042}, Akshara~Viswanathan\inst{11}\orcidlink{0000-0002-7507-5985}, HanYuan~Zhang \gkai{张瀚元}\inst{3}\orcidlink{0009-0005-6898-0927}, and Giuseppina Battaglia\inst{15,16}}  
    \institute{School of Astronomy and Space Science, Nanjing University, Nanjing, Jiangsu 210093, China
    \and
    Key Laboratory of Modern Astronomy and Astrophysics, Nanjing University, Ministry of Education, Nanjing 210093, China
    \and
    Institute of Astronomy, University of Cambridge, Madingley Road, Cambridge CB3 0HA, United Kingdom
    \and 
    Universit\'e de Strasbourg, CNRS, Observatoire astronomique de Strasbourg, UMR 7550, 67000 Strasbourg, France
    \and
    Max-Planck-Institut f\"{u}r Astronomie, K\"{o}nigstuhl 17, 69117 Heidelberg, Germany
    \and
    Kapteyn Astronomical Institute, University of Groningen, Landleven 12, 9747 AD Groningen, The Netherlands
    \and
    LIRA, Observatoire de Paris, Universit\'e PSL, Sorbonne Universit\'e, Universit\'e Paris Cit\'e, CY Cergy Paris Universit\'e, CNRS, 92190 Meudon, France
    \and
    Universit\'e C\^ote d'Azur, Observatoire de la C\^ote d'Azur, CNRS, Laboratoire Lagrange, Nice, France
    \and
    Centre national d'\'etudes spatiales (CNES), 2 Place Maurice Quentin, 75039, Paris, France
    \and 
    Centro de Investigaci\'on en Ciencias del Espacio y F\'isica Te\'orica, Universidad Central de Chile, La Serena 1710164, Chile
    \and
    Department of Physics and Astronomy, University of Victoria, P.O. Box 3055, STN CSC, Victoria BC V8W 3P6, Canada 
    \and
    KU-KIST Green School, Korea University
    \and
    Centre for Astrophysics Research, Department of Physics, Astronomy and Mathematics, University of Hertfordshire, Hatfield, AL10 9AB, United-Kingdom
    \and
    Institute of Physics, Laboratory of Astrophysics, Ecole Polytechnique F\'ed\'erale de Lausanne (EPFL), 1290 Sauverny, Switzerland
    \and
    Instituto de Astrof\'isica de Canarias, E-38205 La Laguna, Tenerife, Spain
    \and
    Universidad de La Laguna, Departamento de Astrof\'isica, 38206 La Laguna, Tenerife, Spain 
    % \and
    % The Observatories of the Carnegie Institution for Science, 813 Santa Barbara Street, Pasadena, CA 91101, USA
    % \and
    % N\'ucleo de Astronom\'ia, Facultad de Ingenier\'ia y Ciencias Universidad Diego Portales, Ej\'ercito 441, Santiago, Chile
    }

% Don't change these lines
\date{Received 8 April 2026 / Accepted 29 July 2026}

\abstract
  % Context
   {Tailored narrow-band photometric observations are an efficient way to measure the metallicity of Milky Way stars. The Pristine survey has been mapping the northern sky since 2015 using narrow-band $CaHK$ photometry obtained with the MegaCam imager on the Canada-France-Hawaii Telescope (CFHT). Data Release 1 (DR1) provided a high-precision absolute calibration of the survey observed before October 2022 using \Gaia DR3 XP synthetic photometry, yielding photometric metallicities over $\sim$6,500 deg$^2$, but was limited in depth to the availability of \Gaia spectro-photometry ($G \lesssim 17.65$).}
  % Aims
   {We present the Pristine Data Release 2 (DR2) that includes all Pristine observations until January 2025, corresponding to an additional 1,500\,deg$^2$ of coverage, and including all faint sources down to $G\sim20.5$. This release aims to exploit the full depth of the Pristine observations to uncover the faint metal-poor halo.}
  % Methods
   {We process all images observed from the start of the survey in March 2015 through to January 2025 following the methodology presented in DR1. The core of the calibration utilizes the subset of bright stars with \Gaia synthetic $CaHK_\mathrm{syn}$ magnitudes to train the \texttt{PhotCalib} neural network. We then derive photometric metallicities using the established Pristine ($CaHK, G, G_{\rm BP}, G_{\rm RP}$) to $\FeH$ model.}
  % Results
   {The resulting catalogue covers approximately $\sim$ 8,000 deg$^{2}$ and provides photometric metallicities for 15 millions FGK stars, which show good agreement with recently published, independent Dark Energy Spectroscopic Instrument (DESI) metallicities.}
   %Conclusions
   {With its increased coverage and access to a significantly fainter regime compared to DR1, DR2 significantly increases the volume of the survey. We report more than 200,000 and 6,500 very and extremely metal-poor star candidates, respectively.}

  \keywords{catalogs -- surveys -- stars: abundances -- Galaxy: abundances}

\maketitle

%%%%%%%%%%%%%%%%% BODY OF PAPER %%%%%%%%%%%%%%%%%%
%\linenumbers
\section{Introduction}

The search for low-metallicity stars in the Milky Way is a search for the fossil record of the early Universe; these stars serve as unique probes for understanding the formation and evolution of primordial stellar systems. A continuing effort to find and identify these rare objects dates back to the HK survey \citep{Beers92} and the Hamburg-ESO survey \citep{Christlieb08}. These pioneer surveys demonstrated the success of utilizing prism spectroscopy in the region of the Ca H \& K lines (3900--4000\AA) to efficiently select low-metallicity stars. Over the last couple of decades, this approach has evolved through the use of narrow-band filters covering the same Ca H \& K region, mounted on wide-field cameras to systematically map the sky. Combined with readily accessible optical broadband photometry, the narrow-band photometry provides reliable metallicity estimates for the numerous FGK stars of the Milky Way. Such observations prove more efficient at probing the very metal-poor (VMP, $\feh < -2$) and extremely metal-poor (EMP, $\feh < -3$) regimes than when using only broadband photometry (\ie \citealt{Ivezic2008}, \citealt{Ibata2017}). Two main surveys started building on this narrow- and medium-band photometry strategy a decade ago: the SkyMapper survey, in the south, with its $v$-band filter \citep{Keller2007, Wolf18, Onken19, Chiti21, Onken2024}, and the Pristine survey \citep{Starkenburg2017,Martin2024}. The latter utilizes a custom narrow-band filter centered specifically on the Ca H \& K lines, mounted on the MegaCam imager at the Canada-France-Hawaii Telescope (CFHT) located at Mauna Kea. Such surveys have been shown to be particularly suited at finding the rare stars at the low end of the metallicity distribution function ($\FeH<-2.0$; \eg \citealt{Aguado2019,DaCosta2019}). Other surveys that use similar strategies have since started (e.g., MAGIC \citealt{placco25}), sometimes complemented with additional narrow-band filters that allow for the determination of other stellar parameters (e.g., J-PLUS --\citealt{Cenarro2019}-- or S-PLUS --\citealt{Mendes2019}).

The spectro-photometric data provided in Data Release 3 of \emph{Gaia} for more than 200 million stars (the so-called XP spectra) contain powerful diagnostics of a star's metallicity \citep{Gaia2023}. The data have led to numerous efforts that aim to extract a star's metallicity (and sometimes also stellar parameters such as effective temperature, gravity, elemental abundances, or foreground extinction) from the very low resolution XP spectra \citep[see \eg][]{Rene2023, Zhang2023, Yao2024, Li2024, Xylakis2024, Arentsen2025, Yang2025, Ye2025}. The Pristine survey utilized \emph{Gaia} XP spectra to generate synthetic $CaHK_{\rm syn}$ Pristine-like photometry. Based on that, all the $CaHK$ measurements taken by CFHT/MegaCam are calibrated using \emph{Gaia} synthetic photometry. By combining this narrow-band photometry with the broad-band \emph{Gaia} photometry, metallicities are calculated from a photometric metallicity model that is based on the $CaHK$, $G$, $G_{\rm BP}$, and $G_{\rm RP}$ magnitudes and trained on a large sample of spectroscopic metallicities. \citet[][hereafter \citetalias{Martin2024}]{Martin2024} presented the first public data release (DR1) of the Pristine survey, as well as the Pristine-like photometric metallicities for the more than 200 million stars with XP spectra. A comparison with metallicities from high-resolution spectroscopy shows agreement in the low-metallicity regime down to $\FeH\sim -3.5$. Two metallicity catalogues were published by \citetalias{Martin2024} for stars with \Gaia DR3 XP spectra: an all-sky catalogue calculated from the Gaia-based synthetic $CaHK_\mathrm{syn}$ magnitudes and, when available, the much higher quality photometric metallicities based on the Pristine CFHT/MegaCam observations.

%one based on the Pristine CFHT Ca H \& K photometry within the Pristine footprint, and one based on the all-sky generated \emph{Gaia} synthetic $CaHK$ catalogue, both using the same Pristine photometric metallicity model. In the very bright end, $G\le 13.25$, the stars with Pristine-\emph{Gaia} metallicities below $-2.8$ have 99 $\%$ success-rate to have [Fe/H] $<-2.5$ from high-resolution spectropic follow-up (Gran et al. 2026 in prep.). \zy{success-rate in intermediate and fainter end}

Because DR1 only included stars with XP spectra, it was limited to the bright end of the Pristine survey ($G \lesssim 17.65$). However, the photometric data gathered at CFHT goes much deeper ($G_\mathrm{lim} \sim 20.5$) and can, for instance, reach much further into the Milky Way's outer halo. In this paper, we publish the second data release of Pristine, DR2, which includes all stars observed as part of the survey until the end of January 2025 (compared to the end of March 2024 in DR1). Aside from the increase in footprint and depth, DR2 relies heavily on the processing techniques and the Pristine metallicity model described in \citetalias{Martin2024}, with only minimal changes to the data processing pipeline. The increased depth of DR2 yields significantly increased numbers of very metal-poor (VMP, $\feh < -2$) and extremely metal-poor (EMP, $\feh < -3$) star candidates. 

This paper is organized as follows. We describe the state of the Pristine survey in Section~\ref{sec:pris}. As we follow the same methodology as in \citetalias{Martin2024}, we only briefly describe the data processing and calibration in Section~\ref{sec:calib} and the determination of the photometric metallicities in Section~\ref{sec:met}. The DR2 catalogue is described in Section~\ref{sec:cat}, which also includes external comparisons, and we conclude with a summary in Section~\ref{sec:sum}.

%-------------------------------------------------------------------
\begin{figure*}
    \centering   
    \includegraphics[width=\linewidth]{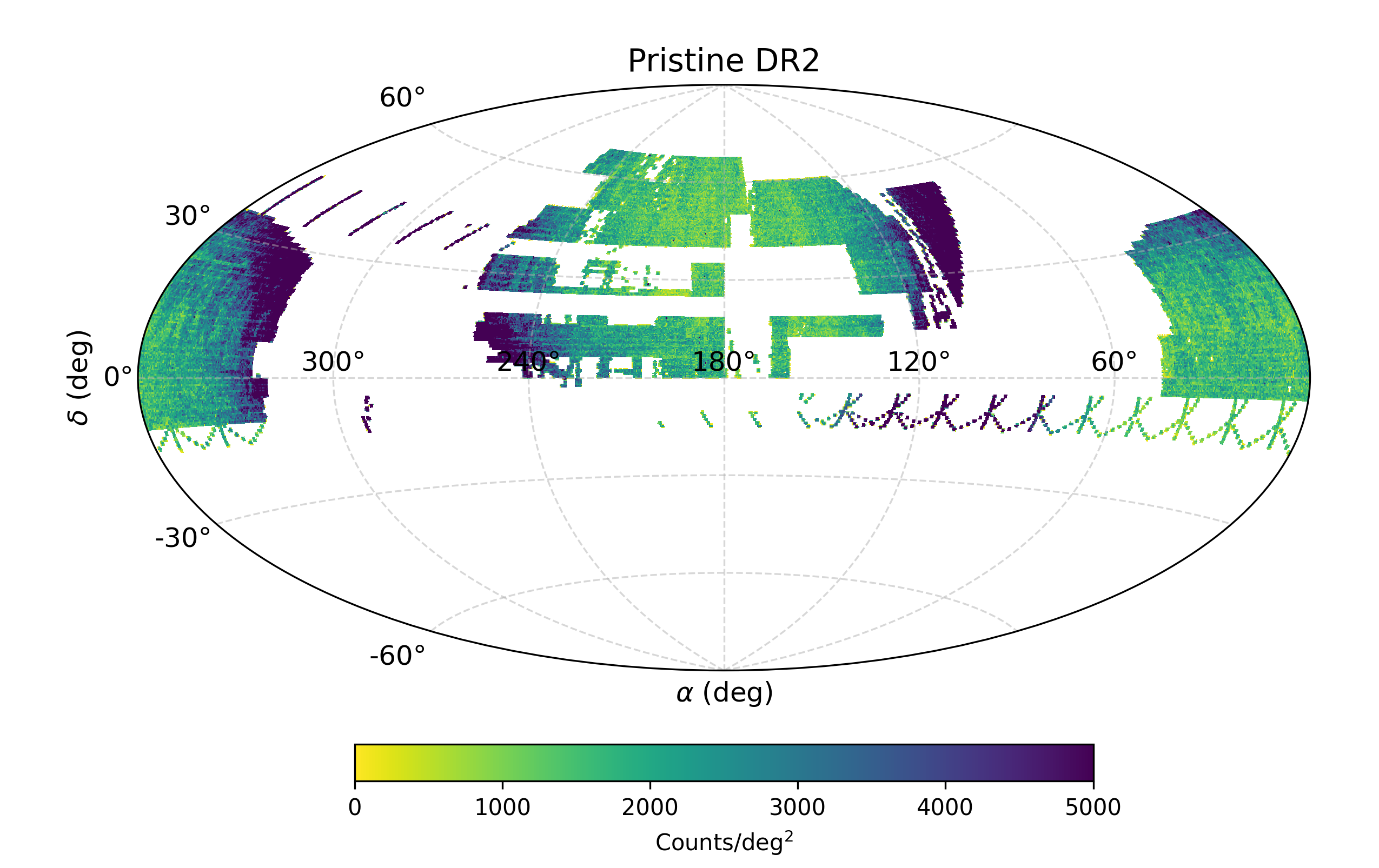}
    \caption{Map of source density in the Pristine DR2 catalogue.}
    \label{fig:skymap}
\end{figure*}
%-------------------------------------------------------------------

\section{Pristine DR2}
\subsection{The Pristine survey}
\label{sec:pris}

The Pristine survey is conducted using the MegaCam wide-field imager on CFHT \citep{Boulade2003} and its narrowband Ca H\&K filter, specifically designed to enable the determination of metallicities for Milky Way FGK stars. Initiated in 2015 from PI programs observed semester after semester, the survey strategy morphed the project into a ``bad-weather'' filler program in 2018, using the MegaCam time that is seldom used because of poor imaging conditions (significant extinction from clouds, low image quality, bright time). As Pristine aims to target relatively bright stars in low-density fields, these observing conditions are mainly sufficient for the science goals. Fields observed under particularly adverse conditions, for which the data reduction can prove difficult, are added back to the observing queue when necessary.

With this survey strategy, Pristine had observed more than 13,000 images by the end of January 2025, which is the date limit for ingestion into DR2 that is the focus of this paper (now 16,900 images in May 2026). These 13,000 images cover $\sim$8,000 deg$^2$, which is $\sim$25\% more than the Pristine DR1 coverage ($\sim$6,500 deg$^2$). The corresponding footprint is shown in Fig.~\ref{fig:skymap}. As $CaHK$ observations are severely impacted by interstellar dust extinction, and the survey's scientific focus is on the metal-poor halo, the survey prioritizes the regions away from the Galactic plane and disk ($|b|\gtrsim25\deg$). The patchiness of the footprint is intrinsically related to the bad-weather strategy but significant effort has been made over the last 2--3 years to fill in as many of the footprint holes as possible. The southern Galactic cap region now includes a large contiguous region of a few thousand square-degrees, in particular because of the poorer weather on Mauna Kea during winter, which was beneficial to Pristine. In the northern Galactic cap, some patchiness remains but will be filled in by a CFHT Large Program that started in 2025B (PI: Martin). This Large Program was awarded time to cover the full contiguous footprint of the WEAVE `Galactic Archaeology Low Resolution high latitude' survey (WEAVE-GA LRhighlat; \citealt{Jin2024}), for which Pristine provides candidate EMP stars for spectroscopic follow-up. Remaining holes in the northern Galactic cap will, as much as possible, be filled through the bad-weather part of the program in coming semesters.

In Fig.~\ref{fig:skymap}, the footprint displays unusual geometric features, below $\delta=-5$\,deg and for $270<\alpha<360$\,deg and $\delta\simeq+35$\,deg. These stem from an unfortunate miscommunication between the Pristine and CFHT teams and the misinterpretation of right ascension coordinates expressed in degrees as expressed in hours. Since these observations are otherwise perfectly valid, they were kept in the survey and reduced like the rest of the data. Some of these fields are, however, located in very high extinction and crowded regions and their data should be used with caution \citep[see][]{Martin2024}.

The Pristine DR2 includes the $CaHK$ photometry and the photometric metallicities of all fields observed until the end of semester 2024B, \ie the data taken until the end of January 2025. DR2 does not include the data taken for the Pristine dwarf galaxy survey \citep[e.g.,\,][]{Longeard2023,Longeard2025}, whose much longer exposures required a different data reduction treatment and will be published in a separate paper (Taibi et al., in preparation). Neither does it include data from the Pristine Inner Galaxy Survey \citep[PIGS; \eg][]{Arentsen2020,Ardern-Arentsen2024} that will be the subject of a specific data release (Ardern-Arentsen et al., in preparation). It is anticipated that the next and final Pristine Data Release, DR3, will be released in 2028, one year after the end of the ongoing CFHT Large Program, scheduled to run until early 2027.

\subsection{Data reduction and calibration}
\label{sec:calib}

A thorough description of the data reduction and calibration of the Pristine images was presented by \citetalias{Martin2024}. Here, we only summarize and detail changes and updates to the procedure since DR1. The data reduction of the Ca H\&K images is performed with the Cambridge Astronomy Survey Unit \citep[CASU;][]{Irwin2001, Irwin2004} pipeline that performs source detection and aperture photometry on the pre-processed images retrieved from the CFHT archive \citep{Magnier2004}. The photometric calibration is performed  using the \texttt{PhotCalib} code\footnote{\url{https://github.com/zyuan-astro/PhotCalib}} \citepalias{Martin2024} to calibrate the $CaHK$ photometry of each image on an absolute scale using the Pristine-like $CaHK_\mathrm{syn}$ photometry generated from the \Gaia DR3 XP information. \texttt{PhotCalib} uses neural networks to determine the field-of-view correction (FOV$(X,Y)$) that depends on the location $(X,Y)$ on the CCD mosaic, as well as the zero point of each individual image by minimizing the residuals between the observed and synthetic magnitudes. FOV$(X,Y)$ is unique to a specific CFHT run, and thus is determined by all the images bundled together for a given run. Both corrections are optimized simultaneously and the best calibration model is generated for each run and image. The resulting calibration model is applied to all stars in the Pristine catalogue, using the specific model of a given image.

A significant portion of the individual images overlap because of repeated observations or because of the overlapping region between fields. Individual detections are merged with their nearby counterparts within 0.5\,arcsec and their magnitudes averaged, using the individual uncertainties as weight. This procedure maximizes the signal-to-noise (S/N) of a source detected multiple times but has the consequence of making the data's depth more inhomogeneous. But the bad-weather strategy followed by Pristine anyway means that the depth of neighboring images may be significantly different. This is visible in Figure~\ref{fig:skymap} that shows variations in the number of sources for nearby regions. As some of the fields included in DR2 partly overlap with the DR1 footprint, the merging procedure has updated the magnitudes of stars in the overlap regions. In other words, while the data have been reduced in the same way as for DR1, the magnitude of a star may change between DR1 and DR2 because it was observed again and its magnitude updated.

The \texttt{merged\_CASU\_flag} tracks both the outcome of the CASU pipeline morphological classifier and the consistency of the merging procedure, with the different values corresponding to:
\begin{itemize}
\item \texttt{-9} for saturated stars for which the photometry is extrapolated at the center of the point-spread function (the Pristine-Gaia synthetic DR1 photometry and metallicity should be used instead for these stars);
\item \texttt{-2} and \texttt{-1} for likely and very likely stars;
\item \texttt{+1} for sources technically identified as extended, but note that this classification is unreliable at the faint end and we recommend including objects with this flag value;
\item \texttt{+2} for merged sources with discrepant individual CASU flags (but these are in most cases still fine and come from a significant number of actual point sources having a CASU flag of \texttt{+1});
\item \texttt{+3} for merged sources with individual $CaHK$ magnitudes discrepant at more than $5\sigma$ between at least two detections.
\end{itemize}

\noindent To isolate likely reliable stars, we use \texttt{-2$\leq$ merged\_CASU\_flag$\leq$+2}.

Finally, all sources are cross-matched with the Gaia DR3 catalogue using a 0.5\,arcsec search radius, keeping the best match. With this procedure, two Pristine detections that are separated by more than 0.5\,arcsec but are both within 0.5\,arcsec of the same \emph{Gaia} source can be assigned to the same \emph{Gaia} source, leading to two sources with the same \emph{Gaia} ID but different Pristine magnitudes and metallicities.

%-------------------------------------------------------------------
\begin{figure*}
    \centering   
    \includegraphics[width=\linewidth]{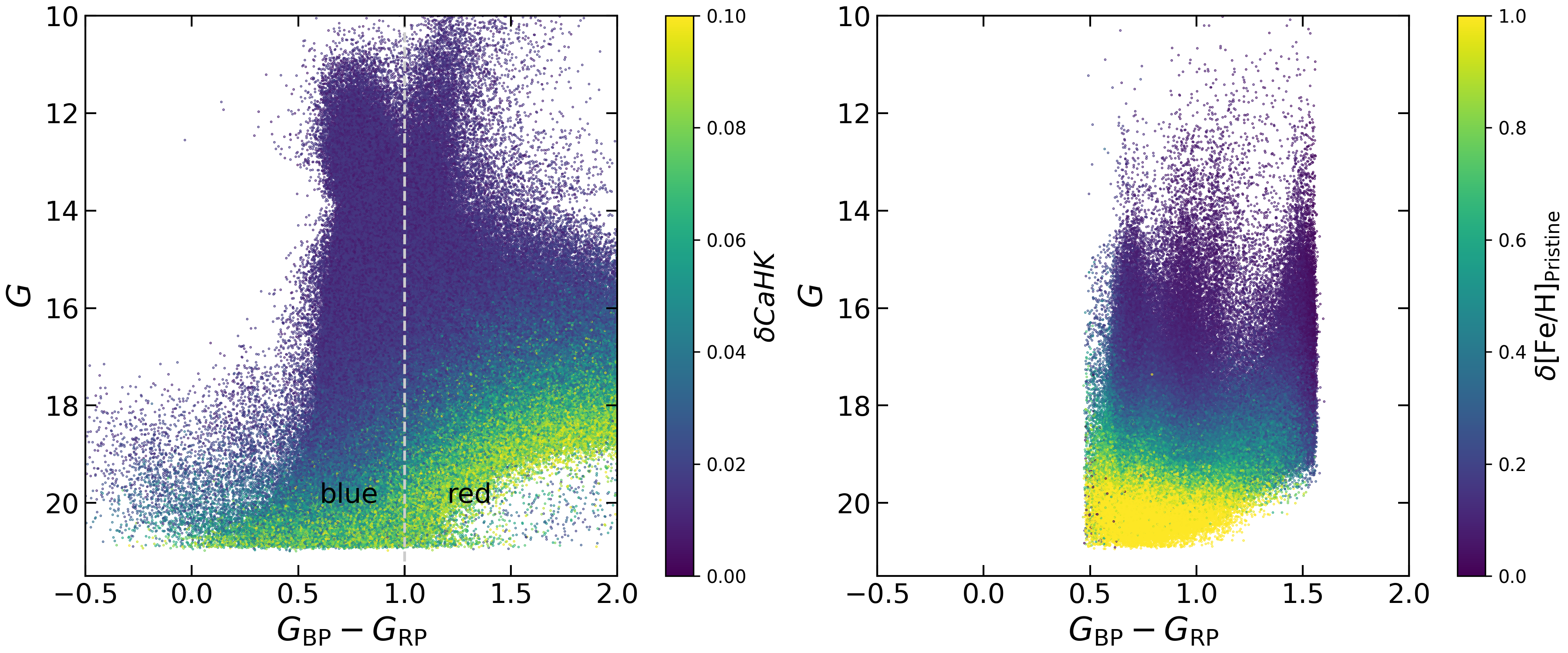}
    \caption{\emph{Left:} Color-magnitude-diagram of the Pristine DR2 stars, color-coded by the uncertainty on the $CaHK$ magnitude, $\delta CaHK$. Stars are selected such that $\delta CaHK \le 0.1$ and following the quality cuts described in Section~\ref{sec:cat}, for a typical halo region ($18\le \alpha \le 210$\,deg, $40 \le \delta \le 60$\,deg). The blue and red stars defined in this work are divided by the gray dashed line at $ G_{\rm BP}-G_{\rm RP}=1$. \emph{Right:} Metal-poor stars ($\FeH<-1.0$) in the same region, color-coded by the uncertainty on the Pristine metallicity, $\delta\FeH_{\rm{Pristine}}$. The color range of stars with metallicities is more restricted as the Pristine photometric metallicity model is only defined for $0.45<(G_{\rm BP}-G_{\rm RP})_0<1.6$. }
    \label{fig:cmd}
\end{figure*}

\begin{figure*}
    \centering   
    \includegraphics[width=\linewidth]{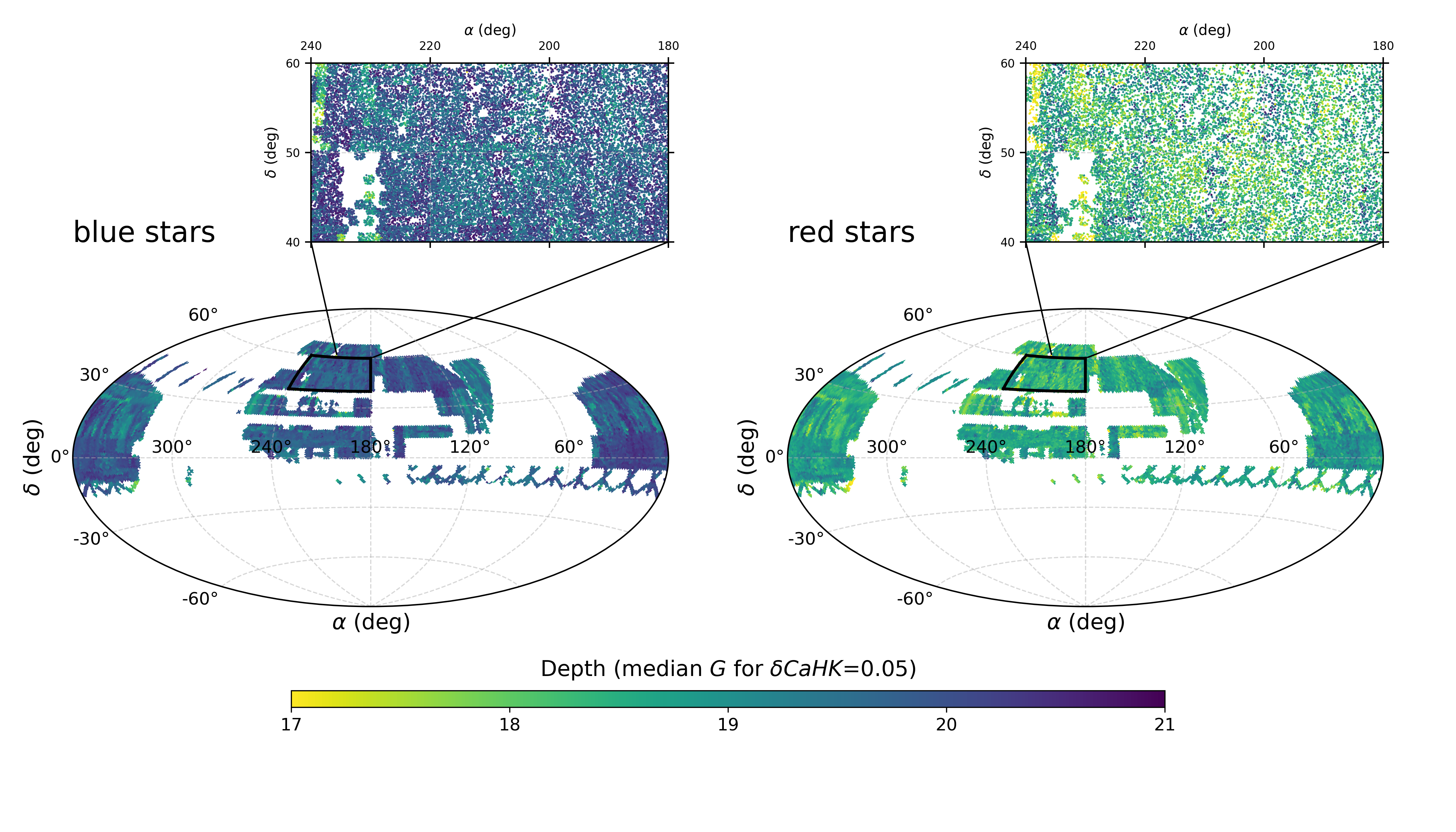}
    \caption{Survey depth as a function of position on the sky for both blue (left) and red stars (right). The survey depth is calculated as the median $G$ magnitude of stars with photometric uncertainties $0.045 \le \delta CaHK \le 0.055$ that also follow the quality cuts listed in Section~\ref{sec:cat}. Because of the blue $CaHK$ filter, the survey is deeper for the blue stars with $0.45\le G_{\rm BP}-G_{\rm RP}\le 1.0$ compared to the red stars with $1.0\le G_{\rm BP}-G_{\rm RP}\le 1.6$. The selected stars color-coded by $G$ are plotted in the insert of each panel and shows the inhomogeneity of a typical halo region.}
    \label{fig:depth}
\end{figure*}
%-------------------------------------------------------------------

The left panel of Fig.~\ref{fig:cmd} displays the color-magnitude-diagram (CMD) of the stars in a typical halo region ($180 \le \alpha \le 210$\,deg, $40 \le \delta \le 60$\,deg) color-coded by the uncertainty in $CaHK$. One can appreciate that the depth of the survey (down to a given $CaHK$ uncertainty) depends on the color of a star. Not surprisingly, at fixed magnitude, blue stars have a higher S/N because the CaHK filter is a very blue filter. At fixed S/N, stars at the turn-off of an old stellar population reach 1.5--2.0 magnitudes fainter than very red stars ($G_{\rm BP}-G_{\rm RP}>1.5$). 

Figure~\ref{fig:depth} shows the survey depth, and its variation, across the survey footprint. Given the strong depth difference between blue and red stars, we show those quantities for both a sample of blue ($0.5<G_{\rm BP}-G_{\rm RP}<1.0$) and red stars ($1.0<G_{\rm BP}-G_{\rm RP}<1.5$), which cover the typical range of stars over which we calculate Pristine photometric metallicities below. We used the magnitude at which $\delta CaHK=0.05$, or equivalently S/N$\sim20$, as a robust tracer of the survey depth ($\delta CaHK=0.1$ or S/N$\sim10$ depths would be $\sim1.5$ fainter). There are strong fluctuations in the measured depths because of the bad-weather strategy followed by Pristine, which leads to variable observing conditions but, overall, the data on average reaches $G\sim19.7$ and $G\sim18.7$ at S/N$\sim20$ for the blue and red stars, respectively (or $G\sim20.3$ and $G\sim19.5$ at S/N$\sim10$).

% Fig.~\ref{fig:uncertainty} shows 1\% of the stars in Pristine DR2 as a representative sample. It is evident that the uncertainty of $CaHK$ increases drastically for stars fainter than $G=18$, same as the results from \citet{Martin2024}. Stars are color coded by $G_{\rm{BP}} - G_{\rm {RP}}$, which show clearly that redder stars have larger uncertainties, resulting from lower $CaHK$ signal-to-noise.

%-------------------------------------------------------------------

% \begin{figure*}
%     \centering   
%     \includegraphics[width=\linewidth]{Pristine_DR2_BP_dCaHK_dFeH.png}
%     \caption{(Left): $CaHK$ magnitude uncertainties for 1\% stars in Pristine DR2, where stars are color-coded by their $G_{\rm BP}-G_{\rm RP}$ color. Redder stars have larger uncertainties in $CaHK$. (Right): uncertainties of photometric metallicity (dwarf solutions) for the same sample of stars. Stars that fainter than $G=19$ and appear bluer have large uncertainties due to the issue with \emph{Gaia} BP photometry.}
%     \label{fig:uncertainty}
% \end{figure*}
\subsection{Photometric metallicities}
\label{sec:met}
\subsubsection{Extinction correction}
\label{sec:extinction}

To correct for the reddening due to interstellar extinction, we implement the same approach as described in detail in \citetalias{Martin2024}. In short, for a given filter $f$, we compute $A_{f} = R_{f} \times E(B-V)$, with $R_{f} = k_{f} \left( \frac{A_{0}}{A_{V}} \right) \times R_V$. In this expression, $A_0$ is the monochromatic extinction at 541.4 nm for which we assume an extinction law with $R_V=3.1$ \citep{Fitzpatrick1999} applied to Kurucz synthetic spectra \citep{CastelliKurucz2003}, and $k_f$ is described as a polynomial function of the stellar parameters ($T_{\rm eff}$, $[\mathrm{Fe/H}]$, $A_0$), using the \texttt{dustapprox} package\footnote{\url{https://mfouesneau.github.io/dustapprox}}. The extinction coefficients in the broad \Gaia filters ($G, G_{\rm BP}, G_{\rm RP}$) depend significantly on the spectral energy distribution (SED) of the source. We derive an effective temperature ($T_{\rm eff}$) using the \cite{Casagrande2021} relation, which mainly depends on the de-reddened \Gaia color and the photometric metallicity from Pristine (detailed below). Since that metallicity depends on color and, therefore, extinction, which itself depends on $T_{\rm eff}$ and metallicity, we iterate this procedure until convergence. We note that the narrow-band Ca H\&K filter is significantly less sensitive to SED variations ($<0.3\%$) than the broadband \Gaia filters. For a typical halo giant, we derived a coefficient $R_{CaHK} \approx 3.92$. The $E(B-V)$ values are taken from the \citet{Schlegel1998} dust maps, updated with the standard recalibration factor of 0.86 from \citet{SchlaflyFinkbeiner2011}. 
%To avoid large systematics introduced by uncertain reddening corrections, we flag or exclude stars in high extinction regions ($E(B-V) > 0.5$) from the high-quality catalogue.

\subsubsection{The photometric metallicity model}
\label{sec:phot_met}

The reader is referred to \citetalias{Martin2024} for the full description of the photometric metallicity model, as well as a thorough assessment of the quality of the derived metallicities. This sub-section only provides a quick summary of the model and resulting metallicities.

We derive photometric metallicities ($[\mathrm{Fe/H}]_{\rm Pristine}$) using the empirically calibrated model presented in \citetalias{Martin2024}. This model maps the location of a star in a de-reddened $(G_{\rm BP}-G_{\rm RP})_0$ vs. $(CaHK-G)_0 - 2.5(G_{\rm BP}-G_{\rm RP})_0$ color-color space to a metallicity value. The training set used to construct the model grid is the same as in \citetalias{Martin2024} and is a combination of spectroscopic metallicities from SDSS/SEGUE \citep{Yanny2009} and APOGEE \citep{Abdurrouf2022}, as well as a handful of high-resolution spectroscopic samples chosen for their size and homogeneity. In order to  boost the training samples for low-metallicity stars with [Fe/H]~$\le -2.0$, the training sample is complemented with stars from the Bo\"otes I dwarf galaxy, a revised version of the LAMOST DR3 VMP sample \citep{Li2018}, as well as Pristine-selected stars confirmed to have $\FeH<-2.0$ during various spectroscopic follow-up campaigns \citep{Youakim2017, Aguado2019, Venn2020,Kielty2021, Lardo2021, Lucchesi2022}. The model grid is also anchored at the lowest metallicities by synthetic colors derived from MARCS (Model Atmospheres in Radiative and Convective Scheme) stellar atmospheres and the Turbo spectrum code \citep{Alvarez98,Gustafsson08,Plez08}. We are aware that there is a slight inconsistency in using extinction coefficients derived from ATLAS models and using MARCS model to anchor the metallicity calibration. However, for stars with effective temperatures above 4000\,K that are the bulk of the stars for which we apply the Pristine calibration the ATLAS and MARCS models are by and large equivalent \citep{Plez2011}.

We apply this model to all sources in the footprint and give two metallicity values for each star, \texttt{FeH\_Pristine} and \texttt{FeH\_Pristine\_dw}, determined from a photometric metallicity model built on the giant and dwarf stars of the training sample, respectively. For most stars, however, it is impossible to know a prior if it is a dwarf or a giant. In the metal-poor regime, for which most stars reside in the more distant Milky Way halo (although see, \eg \citealt{Sestito2020}) we recommend using the model for giants (although see the impact of this choice for a genuine dwarf star in sub-Sect.~\ref{sec:desi}).

More details about the training sample and the generated iso-metallicity lines are shown in Appendix~\ref{sec:app_code}. We introduce a new set of flags, \texttt{Flag\_training} and \texttt{Flag\_training\_dw}. A value of \texttt{0}/\texttt{1} denotes stars distributed above/below the metal-rich edge of the grid, \ie the model of solar metallicity for giants and dwarfs, respectively. This region is poorly sampled by the training set and should be used with caution. We recommend using \texttt{Flag\_training=0} or \texttt{Flag\_training\_dw=0}.

Photometric uncertainties are estimated by performing 100 Monte Carlo realizations, drawing from the magnitude uncertainties for every star. The final catalogue reports the metallicity from the Pristine and Gaia magnitudes (\texttt{FeH\_Pristine} and \texttt{FeH\_Pristine\_dw}), as well as the 16th, 50th and 84th percentiles to enable characterization of the probability distribution function of a star's metallicity. The fraction of Monte Carlo samples that falls within the limits of the metallicity grids are denoted \texttt{mcfrac} and \texttt{mcfrac\_dw}.

The CMD in the right-hand panel of Fig.~\ref{fig:cmd} includes the metal-poor stars ($\FeH<-1.0$) in the same halo region used to build the CMD in the left-hand panel but, this time, color-coded by the uncertainty on the Pristine metallicity for the giant solution, 0.5(\texttt{FeH\_Pristine\_84th}$-$\texttt{FeH\_Pristine\_16th}). The color range is more restricted as the Pristine photometric metallicity model is only defined for $0.45<(G_{\rm BP}-G_{\rm RP})_0<1.6$.

Note that the implementation of the computation of Pristine photometric metallicities\footnote{\url{https://github.com/ankearentsen/Pristine_code}} has been slightly updated. We have switched from using the \texttt{vaex} package to the \texttt{polars} package as the tool to deal with large tables, and we have improved the efficiency of the code. The interpolation within the metallicity grid is slightly different, sometimes resulting in small differences in the photometric metallicities derived for the same input photometry. This has minimal impact for the bulk of the stars, with somewhat larger effects at the metal-poor and blue boundary of the metallicity grid. This is described and shown in Appendix~\ref{sec:app_code}. As we have mentioned in Sect.~\ref{sec:calib}, because the \CaHK magnitudes may have changed between a star's entry in DR1 and DR2 due to more observations contributing to the merged magnitude, a star can also have different photometric metallicities between DR1 and DR2.

\subsection{The DR2 catalogue}
\label{sec:cat}
The Pristine DR2 catalogue contains $\sim$ 22 million sources, $\sim15$ of which have photometric metallicities. It includes all stars from DR1 plus the fainter populations previously excluded, as well as an increase in coverage. Table~\ref{tab:dr2_columns} lists the columns included in this data release. Compared to DR1, we add two more columns (\texttt{Flag\_training} and \texttt{Flag\_training\_dw}) to flag stars that are distributed below the solar metallicity line of the model in the color-color space (see Fig.~\ref{fig:color}). As a best practice, we recommend catalogue users apply quality criteria similar to those presented in \citetalias{Martin2024}. These include highly recommended quality cuts to remove likely spurious metallicities, along with more flexible cuts that can be tailored to the science question explored with the data.

We highly recommend applying the following quality cuts:
\begin{itemize}
\item \texttt{FeH\_Pristine}$<0.0$ and \texttt{FeH\_Pristine}$>-3.9999$\footnote{This cut aims at removing stars with exactly $\FeH_\textrm{Pristine}=-4.0$, but is set to $-3.9999$ instead to avoid rounding errors when reading the catalogue.} to remove stars located at the edge of the metallicity grid. However, for the search of the most metal-poor stars, stars of very low true metallicities may be at the edge of the grid because of photometric uncertainties (see \citepalias{Martin2024});
\item $-2 \le$ \texttt{merged\_CASU\_flag} $\le2$ to avoid saturated stars ($-9$) and sources that have conflicting flags during the merging of their individual detections (+3);
\item \texttt{Flag\_training} = 0 to exclude stars located in the metal-rich region of the color-color space that is not well covered by the training sample.
\end{itemize}

We recommend also using the following criteria, which can be somewhat relaxed depending on the science case:
\begin{itemize}
\item RUWE (renormalised unit weight error) $<$ 1.4 \citep{Lindegren2021} and $|C^*| < 3\times \sigma_{C^*}$, to follow \emph{Gaia} guidelines and ensure the stars have consistent \emph{Gaia} colors and magnitudes (see \citealt{Riello2021} and, in particular, their equations~6 and 18).
\item $P_\mathrm{var}<0.3$ to remove stars that are likely variable, where $P_\mathrm{var}$ is defined as in \citetalias{Martin2024} and conveys the excess of photometric variability of a star;
\item \texttt{mcfrac\_Pristine}$>$0.8 to ensure that the fraction of the magnitude Monte Carlo samplings generally stay within the bounds of the metallicity grid; 
\item $\delta\FeH=0.5(\texttt{FeH\_Pristine\_84th}-\texttt{FeH\_Pristine\_16th})<0.3$ to remove stars with large metallicity uncertainties. However, the metallicity uncertainties tend to increase significantly at the metal-poor end. A science case focusing on those stars may instead preferentially use $\delta\FeH<0.5$;
\item $E(B-V)<0.3$ to avoid highly extincted regions.
\end{itemize}

In what follows, we apply the quality cuts from the first group and we usually apply the cuts from the second group. We clearly specify it when we relax some of these cuts (e.g.\,on the photometric uncertainties). We also note that, as emphasized in \citetalias{Martin2024}, the Pristine metallicity model has not been designed to perform well in the regime [Fe/H]$>-1.0$. While it performs reasonably for such stars, the impact of a star's gravity on its observed $CaHK$ magnitude renders this metallicity range particularly difficult to explore with $CaHK$-based photometric metallicities. Additionally, [$\alpha$/Fe] variations can play a role.

\section{Results}
\subsection{Numbers of VMP and EMP star candidates}

\begin{figure*}
    \centering   
    \includegraphics[width=\linewidth]{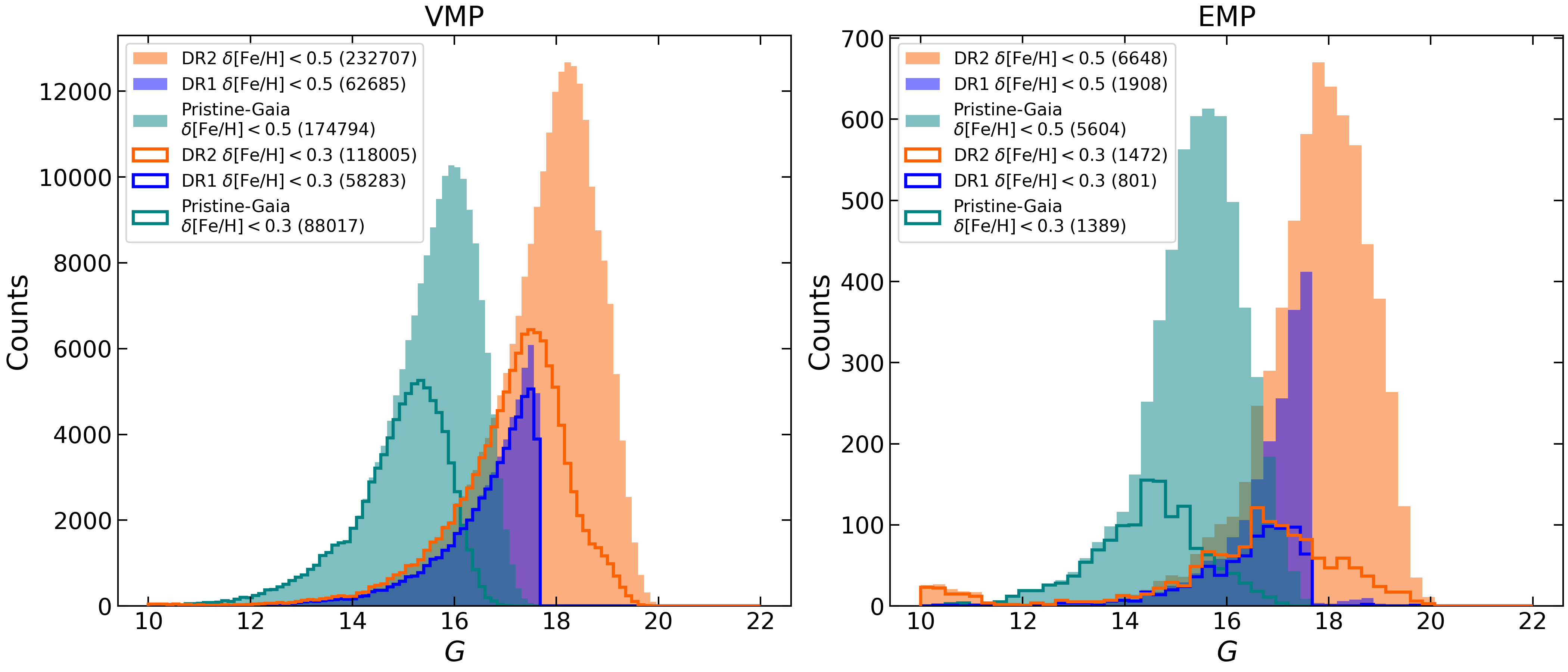}
    \caption{$G$ magnitude distribution of VMP (left) and EMP (right) star candidates for the Pristine DR2 (orange; this paper), DR1 (blue; \citealt{Martin2024}), and the Pristine-\Gaia synthetic data based on Gaia XP (green; \citealt{Martin2024}). Thick lines show the distributions when assuming $\delta\FeH<0.3$ while the full histograms represent counts for $\delta\FeH<0.5$ (all other quality cuts listed in Sect.~\ref{sec:cat} are applied).}
    \label{fig:hist}
\end{figure*}

One of the key science goals of Pristine is the search for stars in the low metallicity regime. Past searches based on the Pristine photometry have proven very successful, with success rates of 80--99\% for VMP candidates (\citealt{Youakim2017, Aguado2019, Venn2020, Kielty2021, Lucchesi2022, Caffau2023, Viswanathan2025}, Gran et al. in prep.). We have also shown that the quality cuts described above are critical to limit contamination. For example, variable stars are a major source of contaminants that can be filtered out using $P_\mathrm{var}$ \citep{Bonifacio2019, Lombardo2023}.

After applying the criteria above, there remains a total of 118,005 VMP and 1,472 EMP candidates in DR2 when applying the restrictive $\delta\FeH<0.3$ cut. These numbers become 232,707 and 6,648 for the looser $\delta\FeH<0.5$ cut that enables the inclusion of fainter targets, at the cost of contamination from intrinsically more metal-rich stars that are pushed into the list of candidates because of the larger uncertainties allowed.

These candidate numbers of both VMP and EMP stars are 2--4 times larger than those reported in \citetalias{Martin2024}. As shown in Fig.~\ref{fig:hist}, this increase is partly due to the increased footprint of DR2 compared to DR1 but, also, mainly to the inclusion of the faint end of the Pristine photometry in DR2 (compare the orange histograms with the blue ones). Despite being significantly shallower than DR2, the Pristine-\Gaia synthetic catalogue, presented in \citetalias{Martin2024}, contains a commensurate number of VMP and EMP candidates compared to DR2 because the former covers the whole sky. However, the DR2 candidates clearly sample a much fainter regime than the Pristine-\Gaia synthetic catalogue enabling the study of the metal-poor Milky Way at much larger distances.

\subsection{Comparison with DESI DR1} 
\label{sec:desi}
\citetalias{Martin2024} showed the very good agreement between the Pristine photometric metallicities and various spectroscopic data sets. The comparisons showed that the Pristine metallicities are accurate at the 0.2--0.3 level. The first large spectroscopic survey of stars in the Northern hemisphere that includes faint stars is the Sloan Digital Sky Survey (SDSS), followed by its successive iterations \citep{Yanny2009, Smee2013}. As we used this survey's spectroscopic metallicities in our training sample to build the Pristine photometric metallicity model, they do not provide a truly independent comparison for the faint Pristine stars, although we did use them to internally check the consistency of our model (further independently confirmed by the results of our own spectroscopic follow-up campaigns). Recently, the Dark Energy Spectroscopic Instrument issued its first data release \citep{DESI_EDR_2024} that includes spectroscopic metallicities for a large number of stars down to $r\sim19.0$ \citep{Cooper2023, Koposov2026} and offers an additional comparison sample for the Pristine photometric metallicities.

We adopt the DESI stellar parameters ($\log g$, [Fe/H]) from the {\tt RVS} pipeline \citep{Koposov2024}, with the metallicity corrections described by \citet{Koposov2026}, and isolate stars with the following cuts recommended in these two papers:

\begin{enumerate}
    \item $\texttt{RR\_SPECTYPE} = \texttt{STAR}$;
    \item $\texttt{RVS\_WARN} = 0$;
    \item $\texttt{snr} > 10$.
\end{enumerate}

\noindent We then cross-match the resulting set of DESI DR1 stars with Pristine DR2. We use DESI $\log g<$3.9 to build a sample of giant stars (5.6\% of the stars from the cross-matched sample of 760,292 stars), with the rest assumed to be dwarf stars ($\log g\ge$3.9). 

\begin{figure*}
    \centering   
    \includegraphics[width=\linewidth]{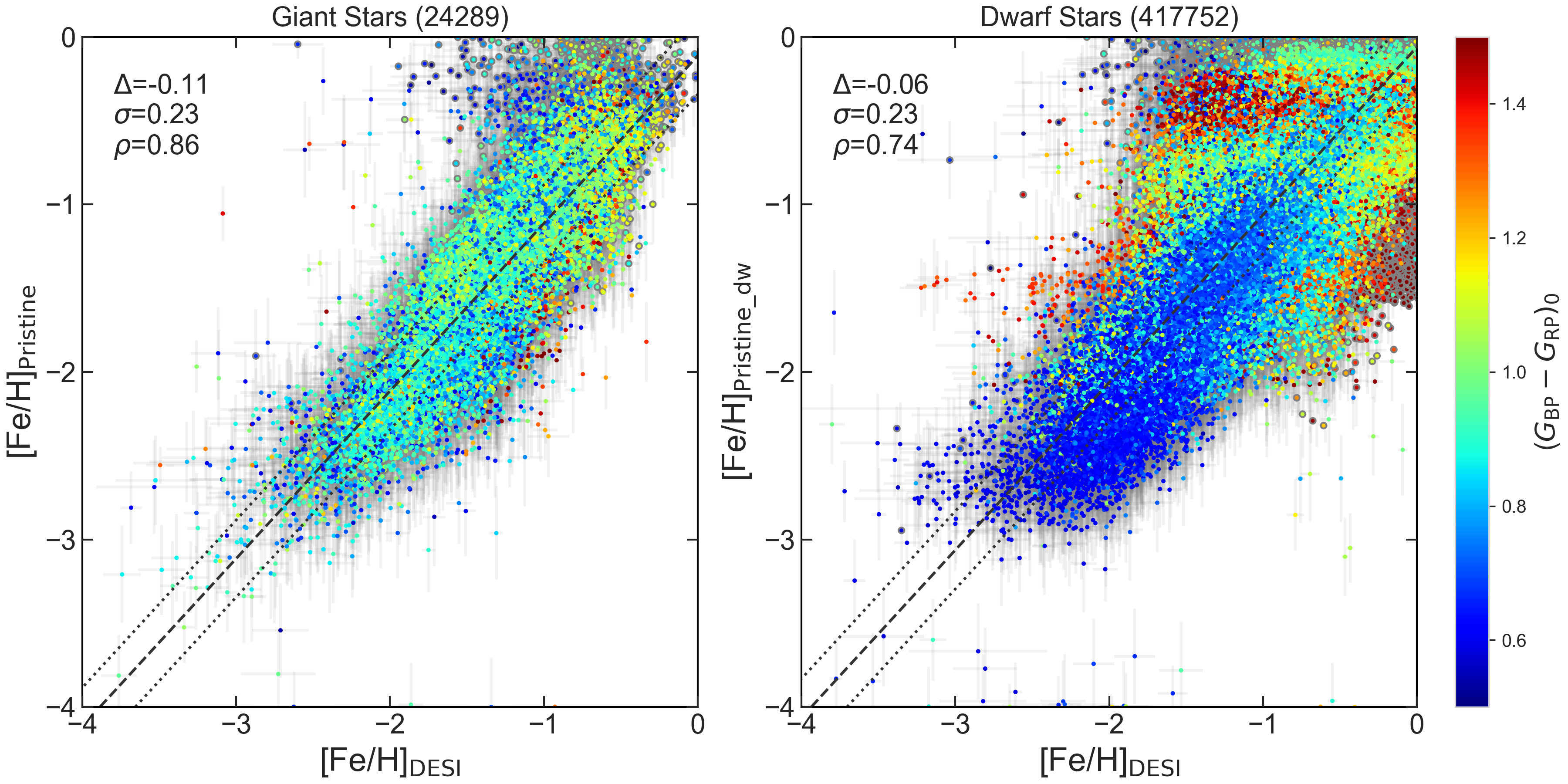}
    \caption{Comparison of metallicities from Pristine DR2 and DESI (corrected following \citealt{Koposov2026}) for giant (left) and dwarf (right) stars. The Pristine quality cuts described in the text are applied and the giant and dwarf classification is based on the $\log g$ values from the DESI RVS pipeline \citep{Koposov2024}. Stars are color-coded by $(G_\mathrm{BP}-G_\mathrm{RP})_0$ as a proxy for temperature. Stars flagged with \texttt{Flag\_training=1} or \texttt{Flag\_training\_dw=1}, distributed below the solar-metallicity line in the color-color space, are further highlighted with gray circles. They mainly contribute to the metal-rich populations in both surveys. In each panel, the dashed line represent the mean one-to-one relation between the two data sets, offset by $\Delta$, as reported in the label, and the dotted line represent the dispersion of the differences ($\pm\sigma$ above/below the dashed line). The correlation coefficient $\rho$ is also reported in the label. More comparisons for the bright ($G \le 17.65$) and faint ($G > 17.65$) sub-samples are shown in the Appendix, in Fig.~\ref{fig:desi2}.}
    \label{fig:desi}
\end{figure*}

Figure~\ref{fig:desi} shows the comparison of the Pristine DR2 photometric metallicities with the DESI spectroscopic metallicities. For giants, shown in the left-hand panel, the two metallicities agree well with each other in the metal-poor regime ($\FeH_\mathrm{Pristine}<-1.0$) for which the Pristine metallicity model is optimized. There is a small systematic offset of $\sim-0.15$\, dex between the two data sets and a dispersion of 0.22 (see the bottom row of Figure~\ref{fig:cmd2} in the Appendix), with the DESI metallicities being higher than their Pristine counterparts for these metal-poor Pristine stars. For dwarf stars, the offset is of $\sim-0.21$ (right-hand panel) and the dispersion is 0.26, slightly larger than for giant stars. A similar offset of roughly 0.2 was previously determined between the corrected DESI metallicities and the metal-poor SAGA (Stellar Abundances for Galactic Archaeology) sample via high-resolution spectroscopy \citep{Koposov2026}.

The Pristine metallicity model performs less well at high metallicity, as revealed by the larger scatter in the metal-rich regime ($\FeH_\mathrm{Pristine} \ge -1$), especially for dwarf stars \citep[see, \eg][]{Caffau2020}. This discrepancy is likely due to the additional impact of $\log g$ on the width of the Ca H \& K lines at higher metallicity. For a typical metal-rich dwarf, the Pristine $CaHK$ magnitudes are therefore a poorer indicator of metallicity. In addition, the behavior of the Pristine metallicities for $\FeH_\mathrm{Pristine} \ge -1$ are impacted by the metallicity limit of the Pristine model that, by design, does not extend beyond $\FeH_\mathrm{Pristine}=0.0$. Any star with a true metallicity above this limit will therefore have an erroneous metallicity in the Pristine catalogue. Finally, some of the deviations observed in the right-hand panel of Fig.~\ref{fig:desi} come from the limited coverage of the training sample in some parts of the Pristine color-color space, particularly at blue end, for $(G_{\rm BP}-G_{\rm RP})_0 \le 0.8$ (see Appendix ~\ref{sec:app_comp} for a more in-depth discussion of this point).

We note a prominent blue population of EMP candidates with $G\approx18$ and $G_{\mathrm BP}-G_{\mathrm RP}\approx0.65$ when including stars with the less strict metallicity uncertainty cut (see Fig.~\ref{fig:cmd2} in Appendix~\ref{sec:app_comp}). This is likely the consequence of a known issue with the \Gaia photometry calibration that causes an overestimation of the mean BP flux for faint sources \citep{Riello2021}. Consequently, some faint sources appear to be more metal-rich than they should be. This effect that is made more problematic by the small distance between iso-metallicity lines for bluer stars.

Overall, the comparison with the DESI metallicities confirms the very good agreement between Pristine photometric metallicities and spectroscopic metallicities at low metallicity, and that, as expected, the Pristine metallicities are less reliable for $\FeH_{\rm Pristine} \ge-1.0$. Appendix ~\ref{sec:app_comp} also details this comparison for different magnitude ranges, brighter and fainter than $G=17.65$.

Finally, with the large sample of DESI dwarf stars, we can explore the consequence of favoring the metallicity model for giants instead of that for dwarfs when the star is a genuine dwarf. Both models agree very well for most stars, with no systematic difference and a dispersion of only 0.07. The distribution however shows a tail with $\sim15$\% of all DESI dwarfs yielding significantly more metal-poor giant metallicities than the dwarf metallicities. This tail is evenly spread between $-1.0<\FeH_\mathrm{Pristine}-\FeH_\mathrm{Pristine,dw}<-0.2$. This emphasizes the importance of using parallax information, for instance from \Gaia, whenever available to choose the most relevant metallicity model.

% \notea{I think we could probably provide a bit more statistics on the comparison. E.g. the median difference and dispersion in multiple metallicity and/or colour/temperature ranges, "contamination" fractions for VMP stars, etc?} \NFM{We feel it's a bit dangerous to push the comparison with DESI too far because it gives the impression DESI is the truth and there's a lot of weird thing with DESI at the metal-poor end that we don't understand. Is the current level of comparison satisfactory?}

\subsection{Globular cluster comparison}

\begin{figure*}
    \centering   
    \includegraphics[width=\linewidth]
    {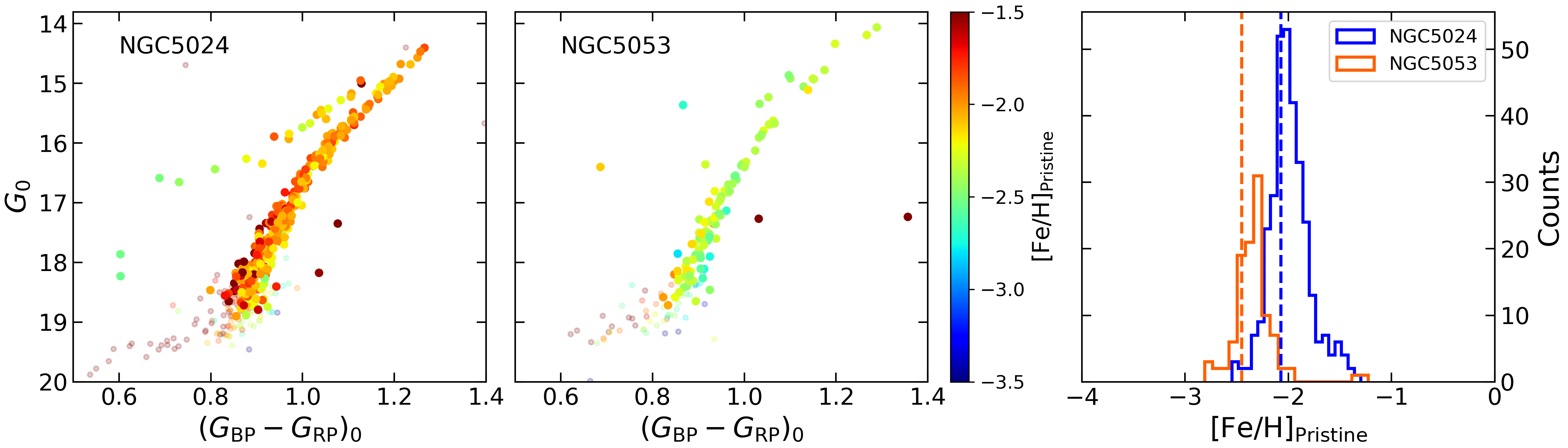}
    \caption{CMDs of the stars around NGC 5024 (left) and NGC 5053 (middle), color-coded by the Pristine metallicities, $\FeH_\mathrm{Pristine}$. Only stars that pass the quality cuts listed in Sect.~\ref{sec:cat} and with metallicity uncertainties $\delta\FeH<0.5$ are shown. Stars with smaller uncertainties ($\delta\FeH<0.3$) are highlighted with larger and more opaque symbols. The right panel shows the MDF of the two clusters for stars with the stricter cut. The peak of the MDF of NGC 5053 is slightly more metal poor than that of NGC 5024, in line with the spectroscopic metallicities of the two clusters shown as dashed lines.}
    \label{fig:gc}
\end{figure*}

Stars in most globular clusters are formed within a relatively short time scale and therefore show only a very small metallicity dispersion. It makes them ideal objects to confirm the reliability of the Pristine photometric metallicities at different magnitudes. A thorough comparison for many clusters was performed as part of the DR1 validation (see sub-Sect.~8.1 and Fig.~21 of \citealt{Martin2024}).

Here, we focus on the very metal-poor globular clusters NGC 5024 (M53) and NGC 5053, which are in the Pristine footprint and have a red giant branch (RGB) that extends over most of the magnitude range of the Pristine data, with a tip of the RGB at $G\sim14.0$--14.5 and a turnoff fainter than the faint end of Pristine DR2. High-resolution spectroscopic studies give $\FeH=-2.07\pm0.07$ \citep{Boberg2016} or $\FeH=-2.06\pm0.13$ \citep{Lamb2015} for NGC 5024, and $\FeH=-2.45\pm0.04$ \citep{Boberg2015} or $\FeH=-2.26\pm0.1$ \citep{Sbordone2015} for NGC 5053.

We select stars around these two GCs in Pristine DR2, following the quality cuts listed at the start of the section, but with the relaxed metallicity uncertainty cut ($\delta\FeH_\mathrm{Pristine}<0.5$). Figure~\ref{fig:gc} shows the CMDs of these two clusters, color-coded by the Pristine metallicities. Since we select all stars around these two GCs in the sky, there is a small number of metal-rich contaminants visible in the CMDs. Pristine provides reliable and consistent metallicities along the red giant branch of the clusters. Deviations only start appearing at the faint end, for $G > 18$, but these are expected due to the photometric uncertainties on $CaHK$ and on $G_\mathrm{BP}-G_\mathrm{RP}$, as can be seen from the broader RGB at fainter magnitude. As a star is observed to be bluer/redder than the true location of the RGB, the color shift will translate into a star that is more/less metal-rich in the Pristine $\left((G_\mathrm{BP}-G_\mathrm{RP})_0, CaHK_0-G_0 - 2.5(G_\mathrm{BP}-G_\mathrm{RP})_0\right)$ color-color space. This is not a failure of the model but is reflected in the larger metallicity uncertainty of those stars, which result from their larger photometric uncertainties. This effect is all the more important for blue stars, whose \CaHK magnitudes are less sensitive to metallicity changes. We note that the metallicities of stars that have entered the helium-core burning phase (left of the RGB) continue to have accurate Pristine metallicities until they become bluer than $(G_\mathrm{BP}-G_\mathrm{RP})_0\sim0.8$.
%reach the horizontal branch phase. There, the spectral features of stars make their metallicity determination unreliable, but Pristine magnitudes, combined with external u-band photometry, can still be useful to isolate those rare stars \citep{Starkenburg2019,Bayer2025}.

The right-hand panel of Fig.~\ref{fig:gc} shows the metallicity distribution functions (MDFs) of the two clusters, as built from Pristine metallicities with the more restrictive metallicity uncertainty cut ($<0.3$). The median metallicities are $\FeH_\mathrm{Pristine}=-2.07$ with a dispersion of 0.15 for NGC 5024, and $\FeH_\mathrm{Pristine}=-2.32$ with a dispersion of 0.13 for NGC 5053. The peak of the MDF agrees well with the high-resolution spectroscopy analyses for NGC 5024 \citep{Lamb2015, Boberg2016}. For NGC 5053, the spectroscopic studies are less consistent but the peak of the Pristine MDF agrees well with the \citet{Boberg2015} value. The small metallicity difference between these two GCs is easily seen in their respective Pristine MDFs.

%-------------------------------------------------------------------

\subsection{Comparison for carbon-enhanced stars}

\begin{figure}
    \centering   
    \includegraphics[width=\linewidth]{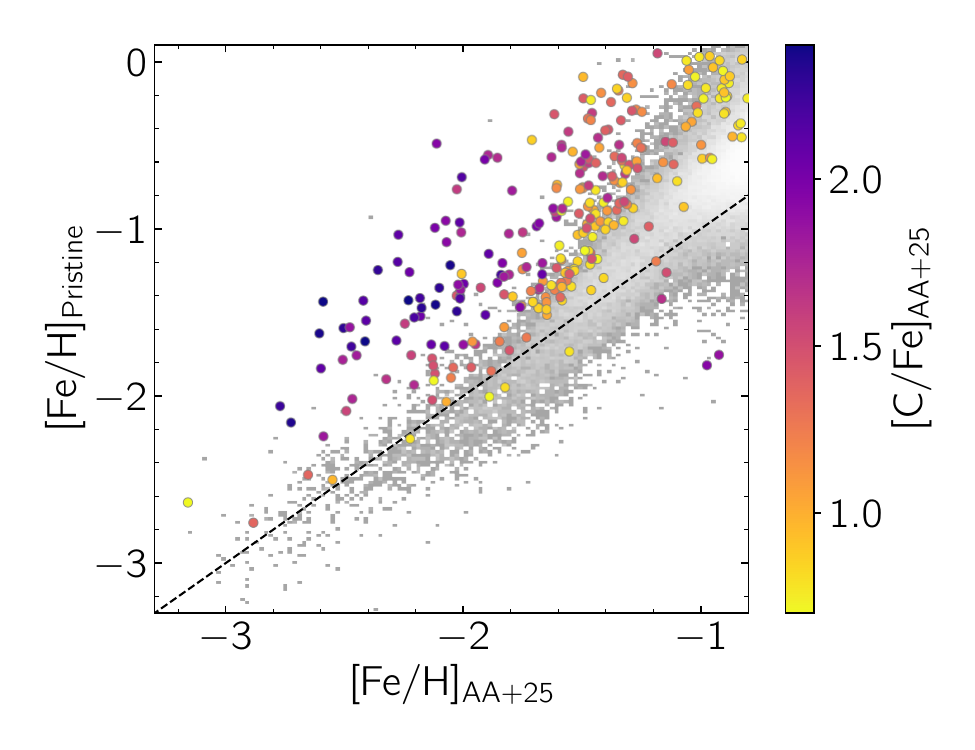}
    \caption{Comparison of Pristine DR2 photometric metallicities with metallicities derived from Gaia XP spectra simultaneously with carbon abundances for a sample of giant stars by \citet[][AA+25]{Arentsen2025}. The background histogram consists of stars with $[\textrm{C}/\textrm{Fe}]~<0.4$ whereas CEMP stars with $[\textrm{C}/\textrm{Fe}]>0.7$ are color-coded according to their carbon abundance. All recommended quality cuts from this work and from \citet{Arentsen2025} have been applied.}
    \label{fig:cemp}
\end{figure}

As discussed in \citetalias{Martin2024} (and references therein), photometric metallicities can be biased for carbon-enhanced metal-poor (CEMP) stars due to their large molecular features that change the color of the stars. CEMP stars are common at low metallicity \citep{Lee13, Placco14} so this bias can have a non-negligible impact on populations of VMP stars derived from photometry. Since the Pristine DR1 paper, new homogeneous samples of CEMP stars have become available. In Fig.~\ref{fig:cemp} we compare the Pristine DR2 photometric metallicities to metallicities determined together with carbon abundances from Gaia XP spectra by \citet{Arentsen2025}. This sample consists purely of cool giant stars ($\sim4700-5300$\,K) for which the effect is expected to be the strongest. Indeed the figure shows that there is a metallicity bias for CEMP stars, which is stronger for more carbon-enhanced stars.

%-------------------------------------------------------------------

% We then further look into these two sample in the color-magnitude-diagram (CMD). The discrepancy between two metallicities arises for both blue and red giants. Especially for the faint and blue stars, it reaches $\sim0.7-1.0\,$dex for both giants and dwarfs, shown in the lower left corner of two panels of Fig.~\ref{fig:desi}. This is cause by the same issue about \Gaia photometry \citep{Riello2021} discussed in Sec.\ref{sec:phot_met}. In short, faint sources G$>$19 appear to be bluer, and also induce problems in determining their metallicities using the photometric metallicity model. 

% \notea{I am also curious about the cross-over between faint stars and higher extinction regions (e.g. $>0.15$) and the amount of contamination there, but not sure we have any way to test this? especially since those regions are typically also closer to the disc = lots of faint disc stars?} \NFM{Do you want to add something about this or do we leave it for DR3?}

% \notea{Anything we can say about previous Pristine papers working with the fainter end of the catalogue, just to emphasize we have already been using it and have done various tests?} \NFM{We added the globular cluster part as a nice way to see how deep Pristine DR2 is. Anything else we can think of?}

\section{Summary}
\label{sec:sum}

In this work, we presented the second data release (DR2) of the Pristine survey. Pristine DR2 extends the survey to a larger footprint with increased depth. DR2 covers $\sim$ 8,000 deg$^2$ of the Northern sky, compared to $\sim$6,500 deg$^2$ for DR1. By calibrating the deep CFHT/MegaCam observations against the absolute reference of \Gaia synthetic photometry, we provide a homogeneous catalogue of 15 million photometric metallicities for stars observed between March 2015 and January 2025, with a limiting magnitude reaching $G\sim20.5$ (compared to $G\sim17.65$ for DR1). 

After applying the quality cuts from Sect.~\ref{sec:cat}, Pristine DR2 contains more than 200,000 VMP, and $\sim$6,500 EMP star candidates with metallicity uncertainties below 0.5, which increases the numbers by 2--4 compared to DR1. These numbers also exceed the brighter but all-sky Pristine-\Gaia synthetic catalogue published as part of DR1. Comparisons with DESI DR1 for candidates fainter than $G=16$ show good agreement, particularly for metal-poor giants that serve as the primary tracers of the Galactic halo. The faintest low-metallicity giant candidates ($\FeH \le -2$, G $\sim$ 20) can probe beyond 100\,kpc, almost tripling the distance that could be reached for stars at the tip of the RGB in DR1. The Pristine EMP candidates from the Pristine DR2 will be targeted in the `WEAVE Galactic Archaeology Low Resolution high latitude' (WEAVE-GA LRhighlat) sub-survey \citep{Jin2024}. These spectroscopic follow-up observations will provide a comprehensive dataset combining 6D phase-space kinematics with chemical abundances for the low-metallicity outer halo. They will allow us to trace the relics from the earliest stellar systems and the building blocks of the Galactic halo.

\section*{Data availability}
The Pristine DR2 is available via \url{https://doi.org/10.5281/zenodo.21092750} or \url{https://sites.google.com/view/pristinesurvey/data-releases}.
\begin{acknowledgements}

We are deeply grateful to the CFHT staff for performing the Pristine observations in queue mode, for their reactivity in adapting the schedule, and for answering our questions during the data-reduction process. 
 
ZY, NFM, and RAI acknowledge funding from the European Research Council (ERC) under the European Unions Horizon 2020 research and innovation programme (grant agreement No. 834148). AA acknowledges support from the Institute of Astronomy's CamCEAD buyout scheme. This work was supported by the ``action th\'ematique'' Cosmology-Galaxies (ATCG) of the CNRS/INSU PN Astro. We benefited from the International Space Science Institute (ISSI) in Bern, CH, thanks to the funding of the teams ``Pristine'' and ``The Early Milky Way''. ZY is supported by the National Key R\&D Program of China via  2024YFA1611601. ES acknowledges funding through VIDI grant "Pushing Galactic Archaeology to its limits" (with project number VI.Vidi.193.093) which is funded by the Dutch Research Council (NWO). This research has been partially funded from a Spinoza award by NWO (SPI 78-411). This work has received funding from the European Research Council (ERC) under the Horizon Europe research and innovation programme (Acronym: EARLYMW, Grant number: 101170507).

Based on observations obtained with MegaPrime/MegaCam, a joint project of CFHT and CEA/DAPNIA, at the Canada-France-Hawaii Telescope (CFHT) which is operated by the National Research Council (NRC) of Canada, the Institut National des Sciences de l'Univers of the Centre National de la Recherche Scientifique of France, and the University of Hawaii. This work has made use of data from the European Space Agency (ESA) mission \Gaia\ (\url{https://www.cosmos.esa.int/gaia}), processed by the \Gaia\ Data Processing and Analysis Consortium (DPAC, \url{https://www.cosmos.esa.int/web/gaia/dpac/consortium}). Funding for the DPAC has been provided by national institutions, in particular the institutions participating in the \Gaia\ Multilateral Agreement. 

Funding for the Sloan Digital Sky Survey IV has been provided by the Alfred P. Sloan Foundation, the U.S. Department of Energy Office of Science, and the Participating Institutions. SDSS-IV acknowledges support and resources from the Center for High Performance Computing  at the University of Utah. The SDSS website is \url{www.sdss.org}. SDSS-IV is managed by the Astrophysical Research Consortium for the Participating Institutions of the SDSS Collaboration including the Brazilian Participation Group, the Carnegie Institution for Science, Carnegie Mellon University, Center for Astrophysics | Harvard \& Smithsonian, the Chilean Participation Group, the French Participation Group, Instituto de Astrof\'isica de Canarias, The Johns Hopkins University, Kavli Institute for the Physics and Mathematics of the Universe (IPMU) / University of Tokyo, the Korean Participation Group, Lawrence Berkeley National Laboratory, Leibniz Institut f\"ur Astrophysik Potsdam (AIP),  Max-Planck-Institut f\"ur Astronomie (MPIA Heidelberg), Max-Planck-Institut f\"ur Astrophysik (MPA Garching), Max-Planck-Institut f\"ur Extraterrestrische Physik (MPE), National Astronomical Observatories of China, New Mexico State University, New York University, University of Notre Dame, Observat\'ario Nacional / MCTI, The Ohio State University, Pennsylvania State University, Shanghai Astronomical Observatory, United Kingdom Participation Group, Universidad Nacional Aut\'onoma de M\'exico, University of Arizona, University of Colorado Boulder, University of Oxford, University of Portsmouth, University of Utah, University of Virginia, University of Washington, University of Wisconsin, Vanderbilt University, and Yale University.
This research made extensive use of the Python programming language
\citep{VanRossum2009} and the following open-source Python packages:
{\tt NumPy} \citep{Harris2020}, {\tt SciPy} \citep{Virtanen2020}, {\tt Matplotlib}
\citep{Hunter2007}, {\tt pandas} \citep{McKinney2010}, {\tt Astropy}
\citep{Astropy2013,Astropy2018,Astropy2022}, and {\tt healpy}
\citep{Zonca2019}.

\end{acknowledgements}



\bibliographystyle{aa}
\bibliography{ms}

%%%%%%%%%%%%%%%%%%%%%%%%%%%%%%%%%%%%%%%%%%%%%%%%%%

%%%%%%%%%%%%%%%%% APPENDICES %%%%%%%%%%%%%%%%%%%%%

\appendix

\section{Updates to the photometric metallicity code}\label{sec:app_code}

\begin{figure*}[!ht]
    \centering   
    \includegraphics[width=\linewidth]{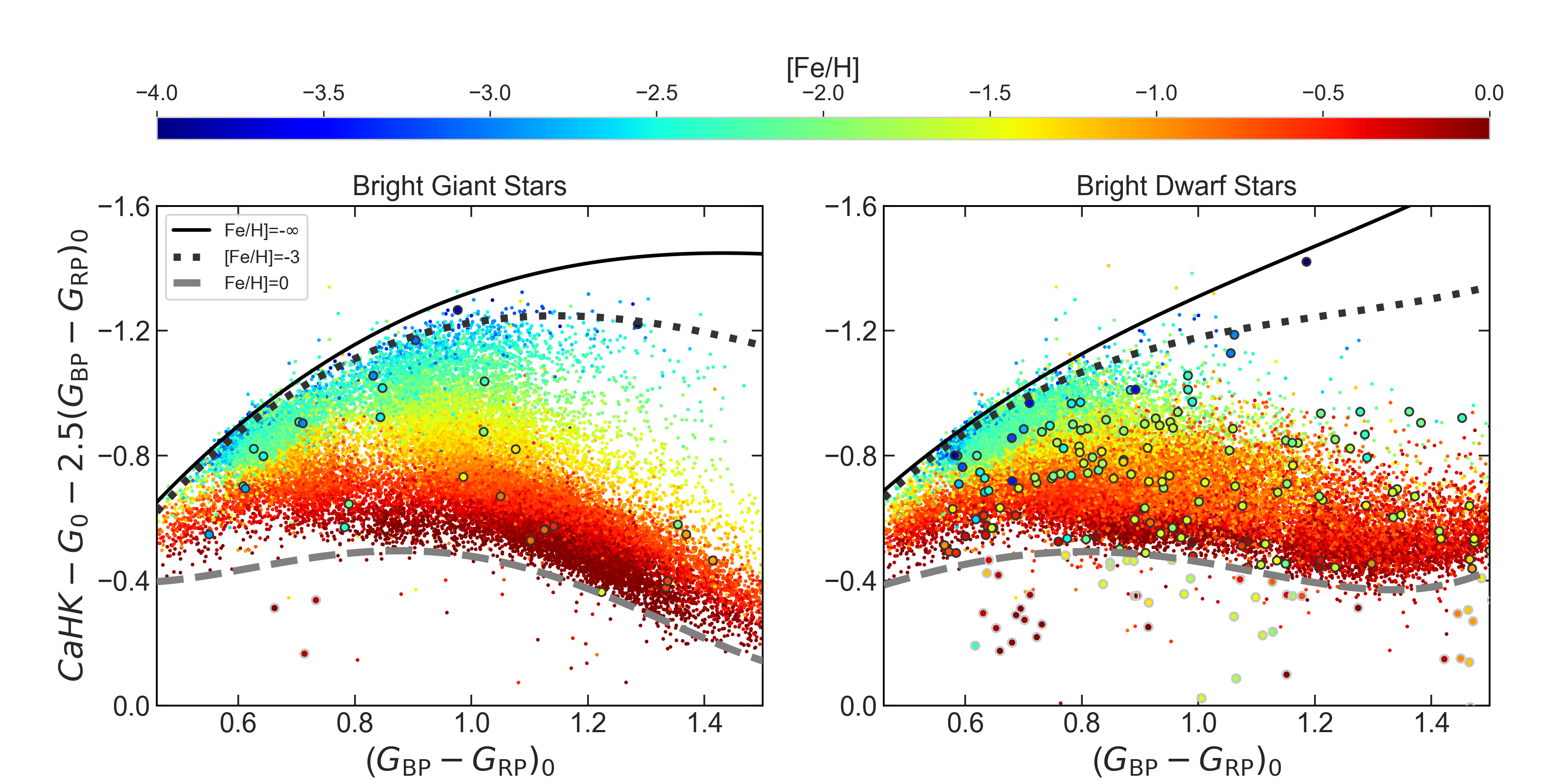}
    \includegraphics[width=\linewidth]    {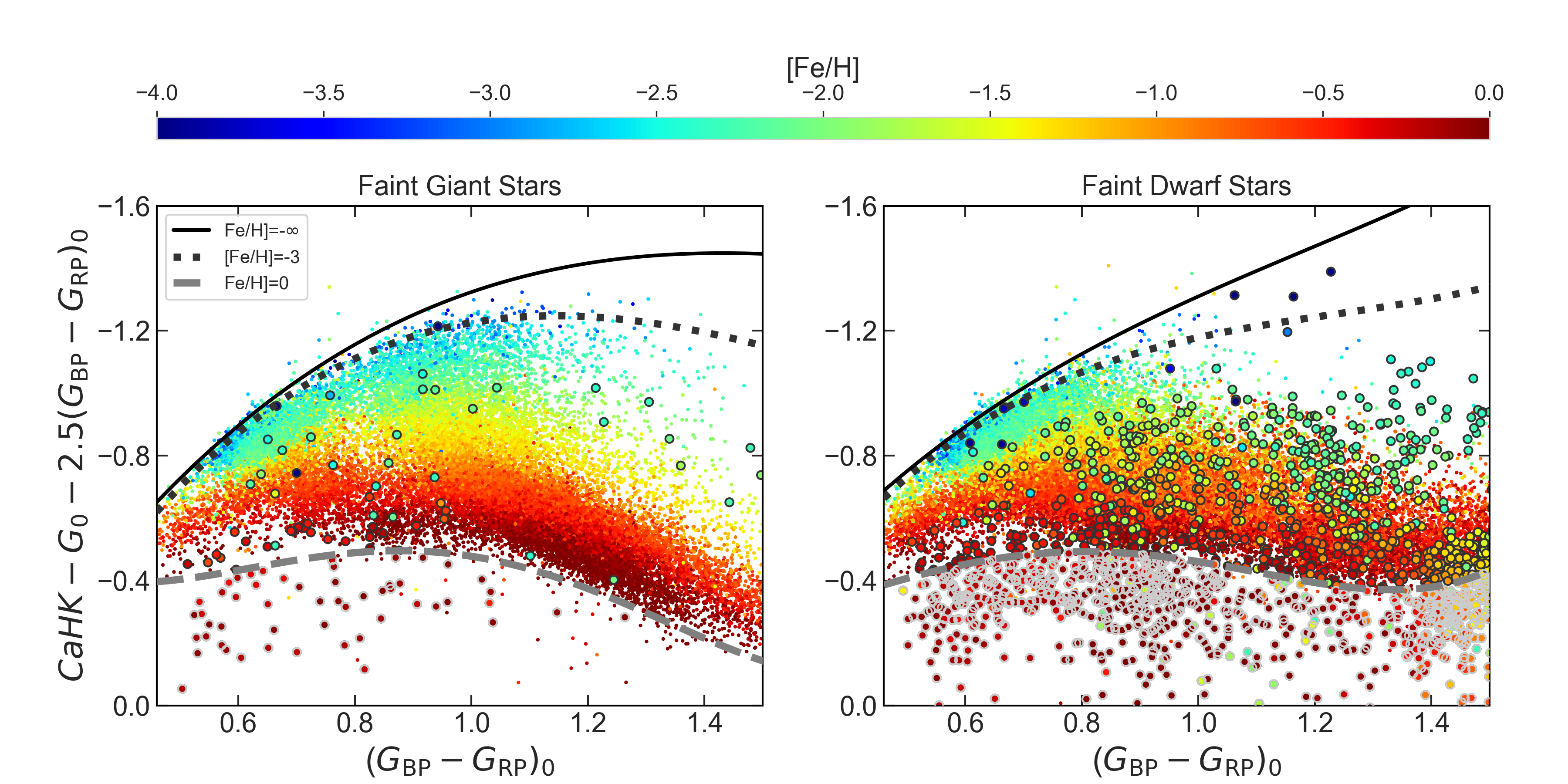}
    \caption{The background shows the photometric metallicity code's training sample in the Pristine color space, color-coded by the spectroscopic metallicities from the literature. Stars in the training sample are mostly distributed between the EMP and solar metallicity lines. The cross-matched DESI stars with large metallicity discrepancy ($\Delta\rm [Fe/H]_{\rm DESI - Pristine} >$1.0 dex) are color-coded by $\rm [Fe/H]_{\rm Pristine}$ and highlighted with black circles if they are above the solar metallicity lines, for bright and faint stars (split by $G=17.65$) in the upper and lower panels, respectively. Those stars below the solar metallicity lines are flagged and plotted with gray circles.}
    \label{fig:color}
\end{figure*}

Figure~\ref{fig:color} shows the training sample in the Pristine color space, color-coded by the metallicities from the literature. Three iso-metallicity lines built from theoretical spectra are overlaid for $\feh = -\infty, -3, 0$ \citep{Starkenburg2017,Martin2024}. Most of the stars in the training sample are distributed between the iso-metallicity lines of $\feh = -3$ and $\feh = 0$.

\begin{figure*}[!ht]
    \centering   
    \includegraphics[width=0.48\linewidth]{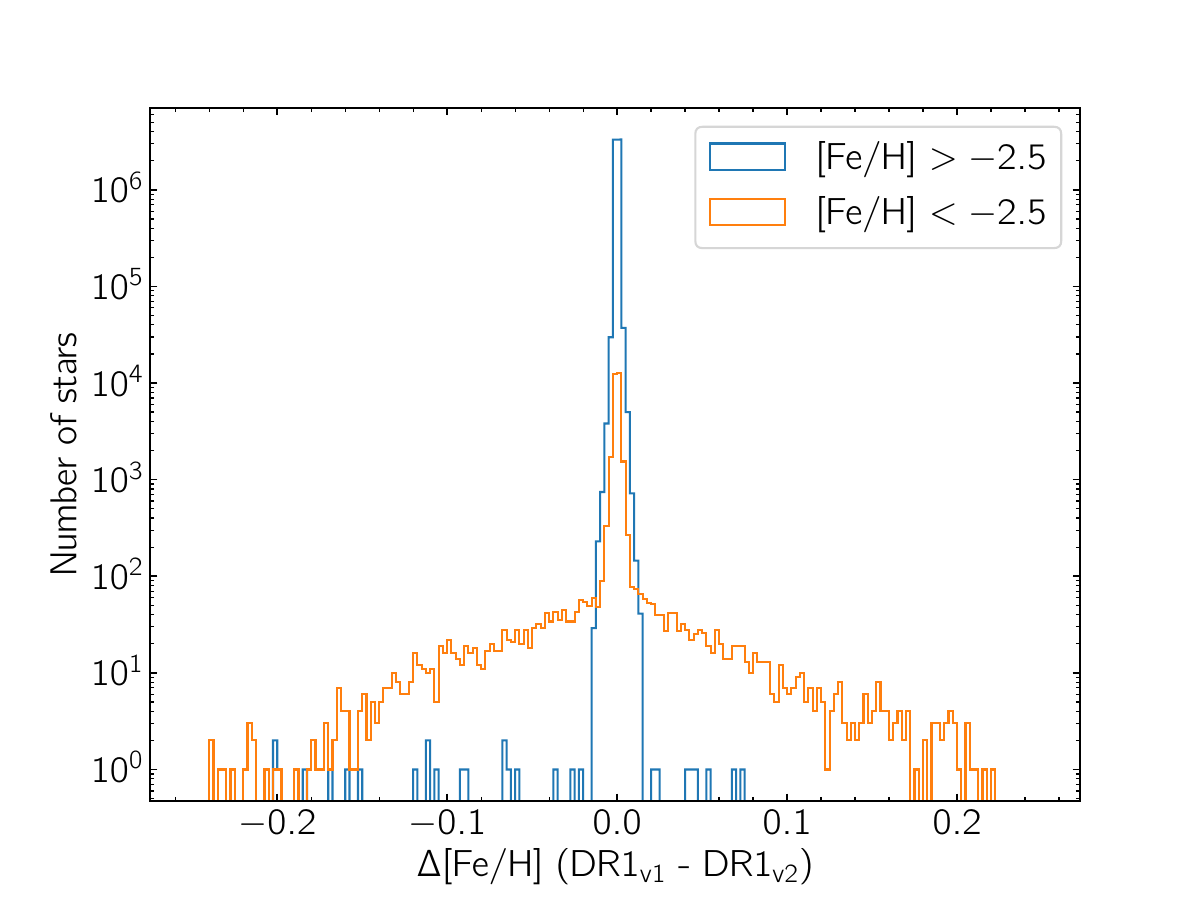}
    \includegraphics[width=0.48\linewidth]{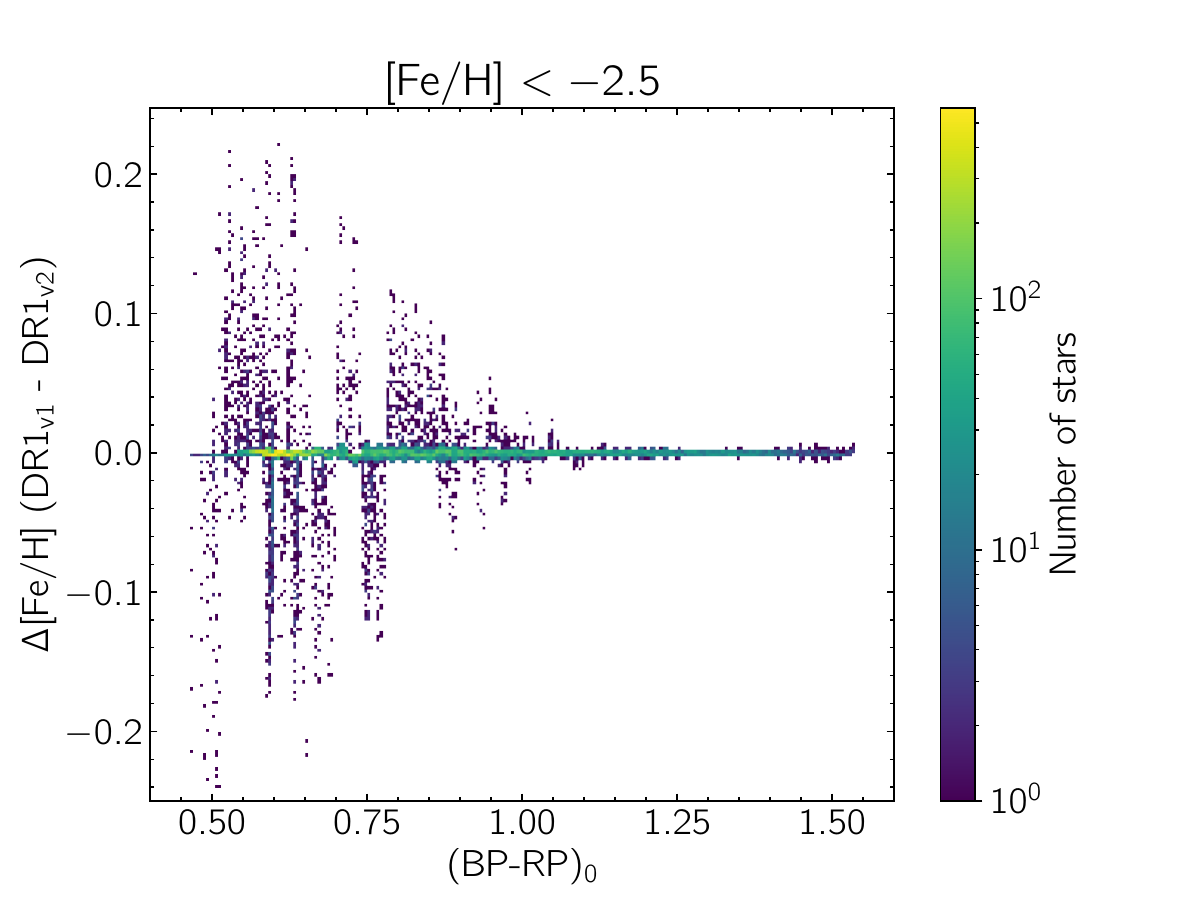}
    \caption{Comparison of giant metallicities from \texttt{v1} \citep{Martin2024} and \texttt{v2} (this work) of the photometric metallicity code, for DR1 data. The left panel shows logarithmic histograms for stars above (blue) and below (orange) $\feh = -2.5$. The right panel shows the metallicity difference for the more metal-poor sample as function of color. Only stars with [Fe/H] uncertainties $<0.3$, E(B$-$V) $< 0.3$ and \texttt{mcfrac} $> 0.8$ are included here.
    }
    \label{fig:polars}
\end{figure*}

In moving the photometric metallicity code from using \texttt{vaex} to \texttt{polars}, we also changed the function used to interpolate within the photometric metallicity grid, from \texttt{scipy.interpolate.griddata} to \texttt{scipy.interpolate.RegularGridInterpolator}. The effect on the metallicities is shown in Fig.~\ref{fig:polars}. It is negligible for stars with $\feh > -2.5$. For $\feh < -2.5$ the differences become noticeable for some stars bluer than $(G_\mathrm{BP}-G_\mathrm{RP})_0< 1.0$. However, these variations only reach a maximum of 0.1--0.2 and are well within the range of typical photometric metallicity uncertainties for warm EMP stars. As can be appreciated from Fig.~\ref{fig:color}, the metallicity variation in this color regime is very rapid.

\section{Description of the Pristine DR2 metallicity catalogue.}

See Table~\ref{tab:dr2_columns}. 

%-------------------------------------------------------------------
\begin{table*}
\caption{Description of the columns of the Pristine DR2 metallicity catalogue.}
\label{tab:dr2_columns}
\centering
\begin{tabular}{l l c c}
\hline\hline
Column & Description & Unit & Type \\
\hline
\texttt{source\_id} & Gaia DR3 \texttt{source\_id} & & longint \\
\texttt{RA} & Gaia DR3 right ascension (ICRS J2016) & degrees & float \\
\texttt{Dec} & Gaia DR3 declination (ICRS J2016) & degrees & float \\
\texttt{E(B-V)} & Schlegel et al. (1998) extinction value & mag & float \\
\texttt{merged\_CASU\_flag} & Morphology and quality flag & & int \\
\texttt{G\_0} & De-reddened Gaia $G$ magnitude & mag & float \\
\texttt{dG} & $\delta G$, uncertainty on the Gaia $G$ magnitude & mag & float \\
\texttt{BP\_0} & De-reddened Gaia $G_{\rm BP}$ magnitude & mag & float \\
\texttt{d\_BP} & $\delta G_{\rm BP}$, uncertainty on the Gaia $G_{\rm BP}$ magnitude & mag & float \\
\texttt{RP\_0} & De-reddened Gaia $G_{\rm RP}$ magnitude & mag & float \\
\texttt{d\_RP} & $\delta G_{\rm RP}$, uncertainty on the Gaia $G_{\rm RP}$ magnitude & mag & float \\
\texttt{CaHK\_0} & De-reddened Pristine CaHK magnitude & mag & float \\
\texttt{CaHK\_Pr} & Pristine CaHK magnitude & mag & float \\
\texttt{d\_CaHK} & $\delta$CaHK, uncertainty on the Pristine CaHK magnitude & mag & float \\
\texttt{RA\_Pr} & Pristine right ascension (ICRS J2015.5) & degrees & float \\
\texttt{Dec\_Pr} & Pristine declination (ICRS J2015.5) & degrees & float \\
\hline
\texttt{FeH\_Pristine} & Photometric metallicity [Fe/H]$_{\rm phot}$ calculated directly from \texttt{CaHK\_0} (giant model) & dex & float \\
\texttt{FeH\_Pristine\_16th} & Lower bound (16th percentile) of [Fe/H]$_{\rm phot}$ PDF & dex & float \\
\texttt{FeH\_Pristine\_50th} & Median value (50th percentile) of [Fe/H]$_{\rm phot}$ PDF & dex & float \\
\texttt{FeH\_Pristine\_84th} & Upper bound (84th percentile) of [Fe/H]$_{\rm phot}$ PDF & dex & float \\
\texttt{mcfrac\_Pristine} & Fraction of MC samples inside model color-color space & & float \\
\texttt{Flag\_training} & 0 if above the zero metallicity isoline for giants, otherwise 1 & & int \\
\texttt{G\_0\_dw} & Same as \texttt{G\_0} but for the dwarf model & mag & float \\
\texttt{BP\_0\_dw} & Same as \texttt{BP\_0} but for the dwarf model & mag & float \\
\texttt{RP\_0\_dw} & Same as \texttt{RP\_0} but for the dwarf model & mag & float \\
\texttt{CaHK\_0\_dw} & Same as \texttt{CaHK\_0} but for the dwarf model & mag & float \\
\texttt{FeH\_Pristine\_dw} & Same as \texttt{FeH\_Pristine} but for the dwarf model & dex & float \\
\texttt{FeH\_Pristine\_dw\_16th} & Same as \texttt{FeH\_Pristine\_16th} but for the dwarf model & dex & float \\
\texttt{FeH\_Pristine\_dw\_50th} & Same as \texttt{FeH\_Pristine\_50th} but for the dwarf model & dex & float \\
\texttt{FeH\_Pristine\_dw\_84th} & Same as \texttt{FeH\_Pristine\_84th} but for the dwarf model & dex & float \\
\texttt{mcfrac\_Pristine\_dw} & Same as \texttt{mcfrac\_Pristine} but for the dwarf model & & float \\
\texttt{Flag\_training\_dw} & 0 if above the zero metallicity isoline for dwarfs, otherwise 1 & & int \\
\hline
\texttt{Pvar} & Probability for the source to be variable & & float \\
\texttt{RUWE} & Gaia DR3 Renormalised Unit Weight Error & & float \\
\texttt{Cstar} & Gaia DR3 corrected flux excess, $C^*$, as defined in equation (6) of \citet{Riello2021} & & float \\
\texttt{Cstar\_1sigma} & Normalized standard deviation of $C^*$, as defined in equation (18) of \citet{Riello2021} & & float \\
\hline
\end{tabular}
\end{table*}
%-------------------------------------------------------------------

\section{Additional comparisons between the Pristine and DESI metallicities}
\label{sec:app_comp}
\begin{figure*}[!ht]
    \centering   
    \includegraphics[width=\linewidth]{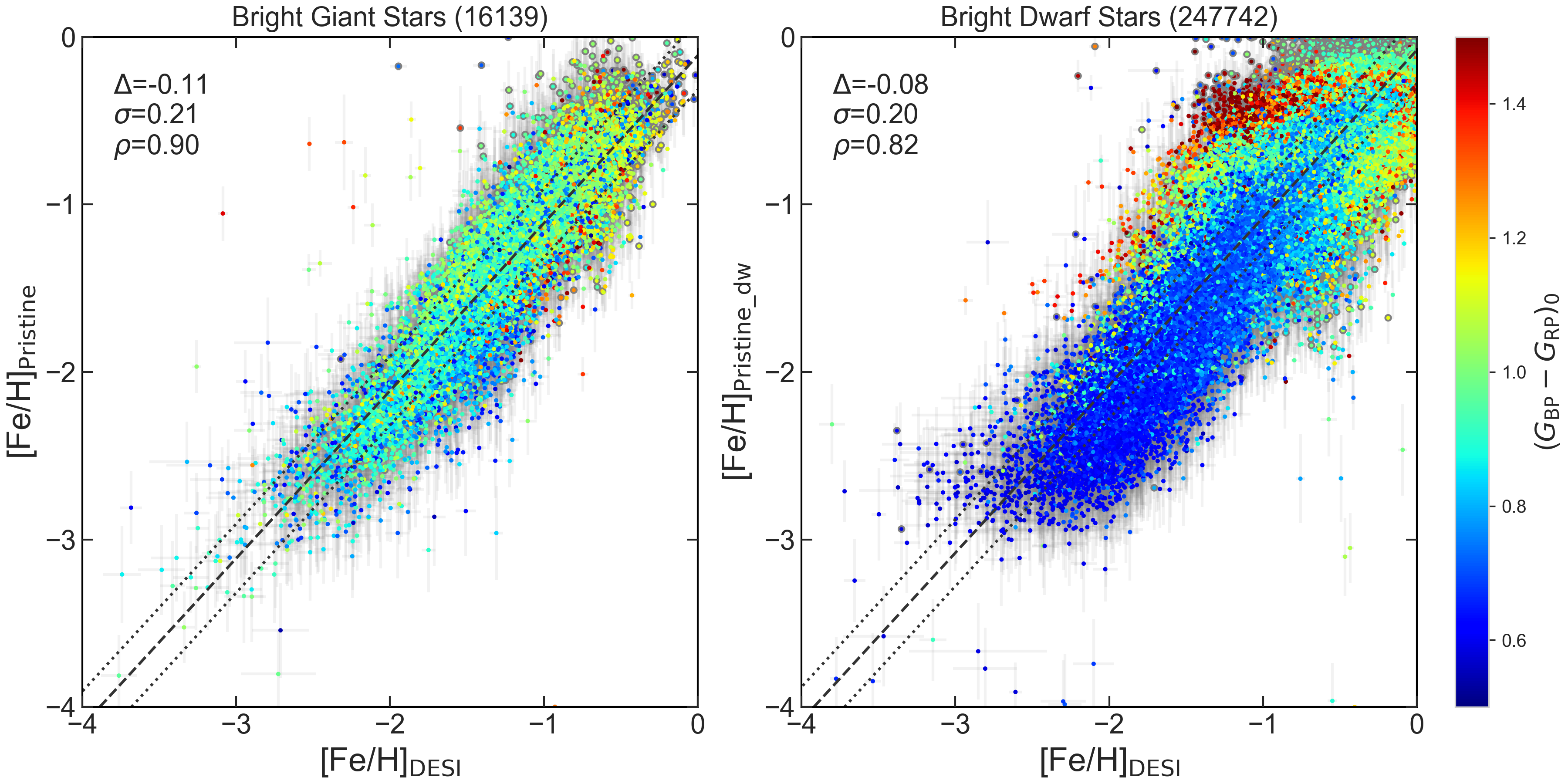}
    \includegraphics[width=\linewidth]{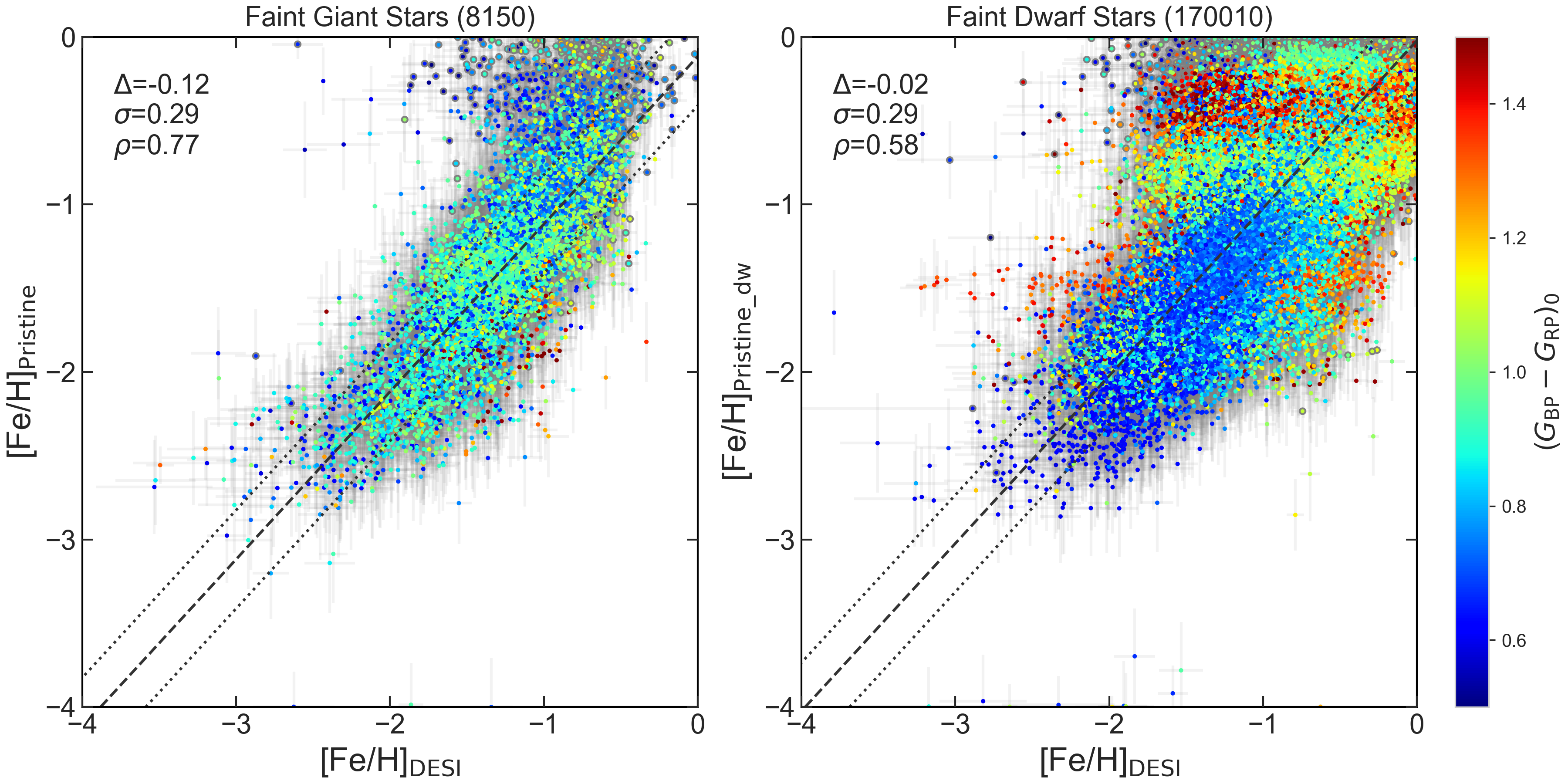}
    \caption{Same as Fig.~\ref{fig:desi}, by with the stars divided into the brighter sample ($G \le 17.65$) and the fainter sample ($G >17.65$). The correlation is excellent for the bright stars and degrades for the faint ones. The flagged stars distributed below the solar-metallicity line in the color-color space are further highlighted with gray circles. These stars are the main contributors of a population of blue ($(G_\mathrm{BP}-G_\mathrm{RP})_0\lesssim 0.8$) giant stars that are more metal rich in Pristine and populate the upper-right corner of the left panels ($\FeH_{\rm Pristine} \ge -1$)}.
    \label{fig:desi2}
\end{figure*}

\begin{figure*}[!ht]
    \centering   
    \includegraphics[width=\linewidth]{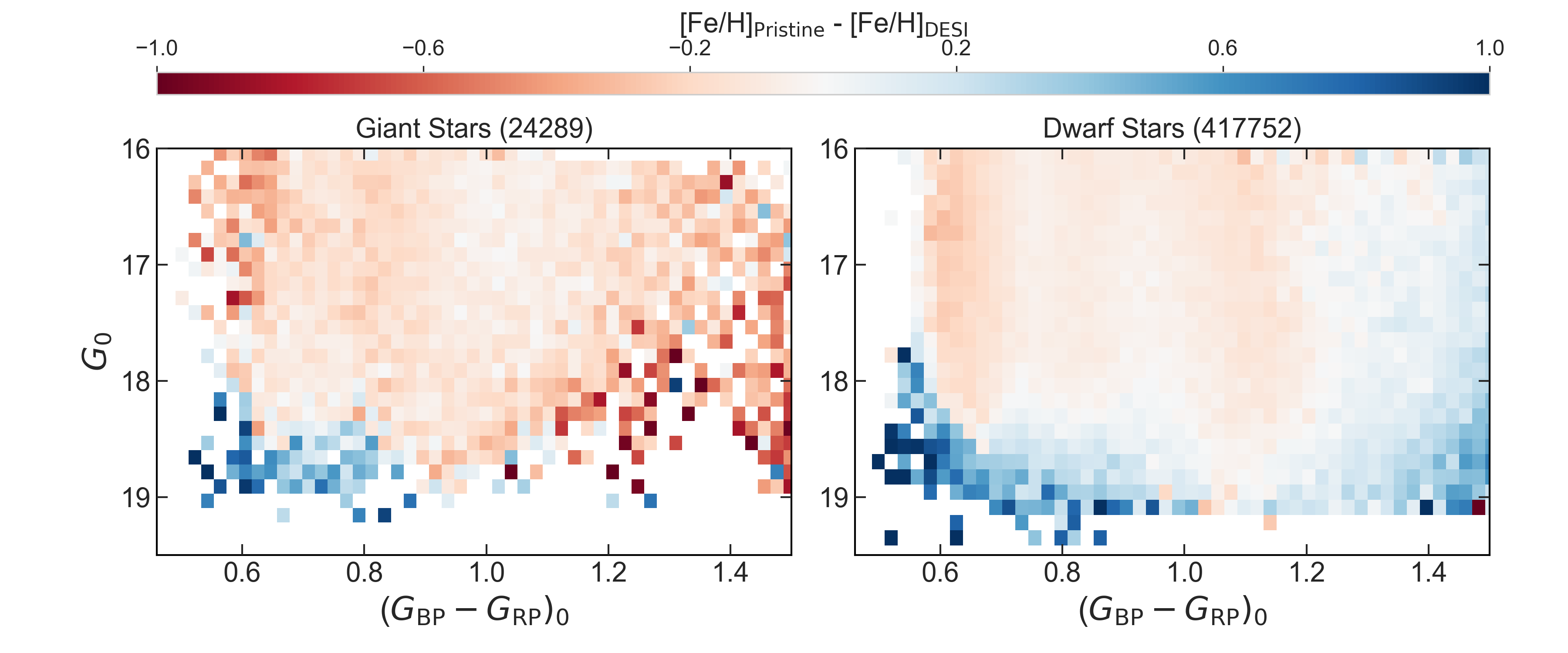}
    \includegraphics[width=\linewidth]{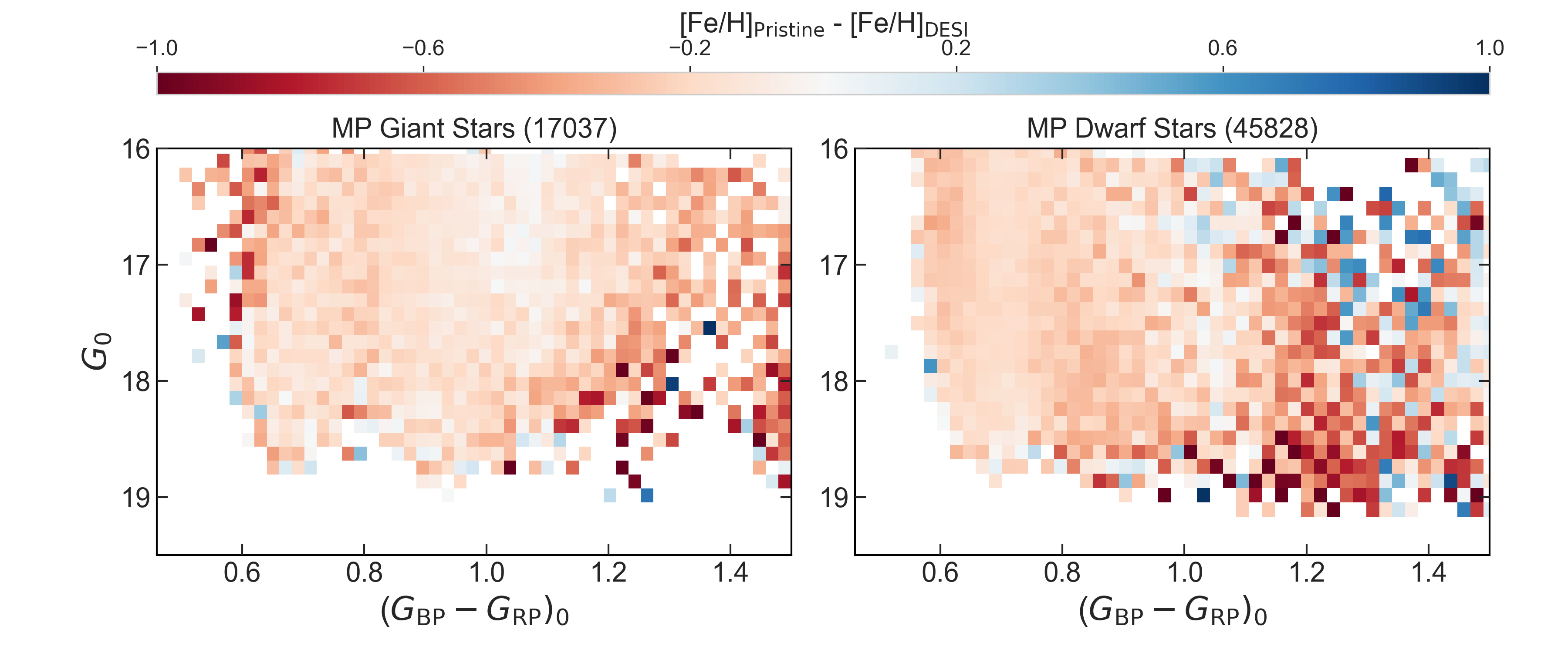}
    \includegraphics[width=\linewidth]{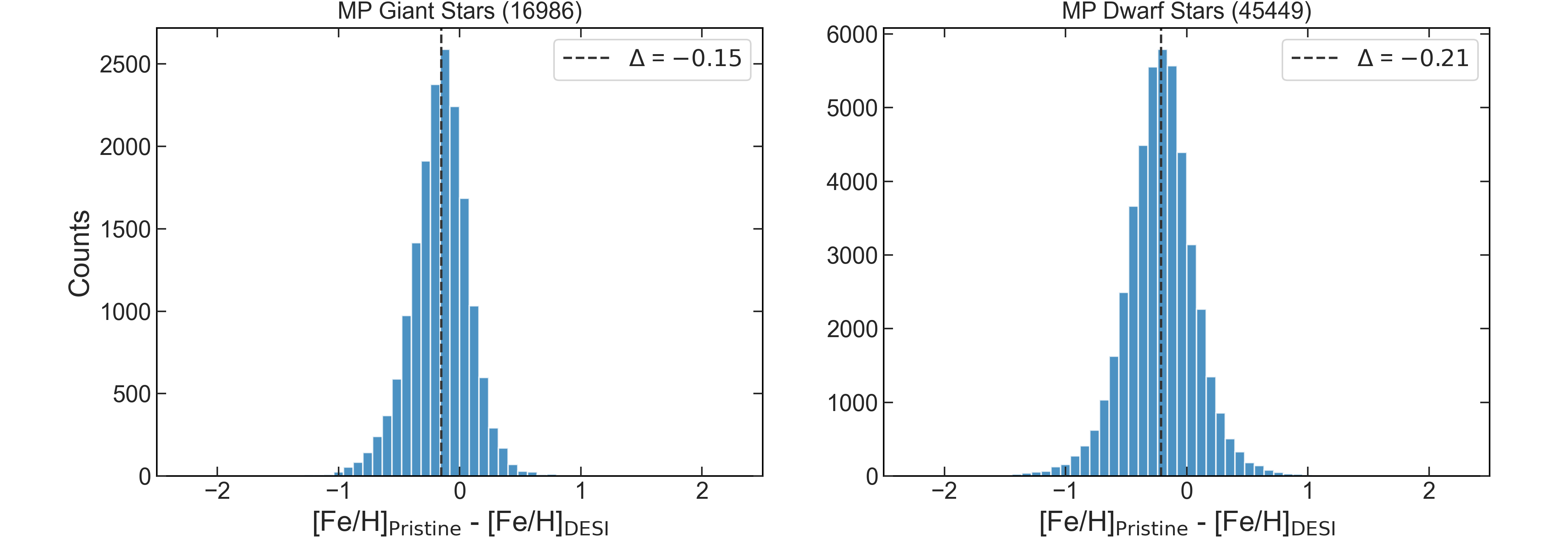}
    \caption{CMD of stars in both Pristine and DESI, color-coded by $\FeH_{\rm Pristine}-\FeH_{\rm DESI}$ (upper row), and CMD of just the metal-poor stars (middle row). The left/right columns show the comparisons for giants/dwarfs, with the classification based on the $\log g$ values from DESI. A population of metal-rich blue stars ($(G_\mathrm{BP}-G_\mathrm{RP})_0 \lesssim 0.8$) shows a systematic metallicity offset between Pristine and DESI in the upper panel (see also Fig.~\ref{fig:desi}). These stars are effectively removed in the lower panel by applying a metallicity cut of $\FeH_{\rm Pristine} < -1$. In the metal-poor regime, DESI has systematically higher metallicities (bottom row). The mean $\FeH_{\rm Pristine}-\FeH_{\rm DESI}$ offsets are $-0.15$ for the MP giants, with a dispersion of 0.22, and $-0.21$ for the MP dwarfs, with a dispersion of 0.26. The means are represented by the black dashed lines.}
    \label{fig:cmd2}
\end{figure*}

Here, we provide some more detailed comparisons of the Pristine and DESI metallicities, complementing the results shown in sub-section~\ref{sec:desi}. In particular, we divide the cross-matched Pristine and DESI sample into two sub-samples using a cut at $G = 17.65$ (the Gaia XP/Pristine DR1 limit). Figure~\ref{fig:desi2} shows the comparisons for the bright (upper) and faint (bottom) samples. The correlation is excellent for the bright stars in the metal-poor regime ($\FeH_\mathrm{Pristine}<-1.0$), and degrades in the faint regime because of the lower S/N of faint stars in both surveys. For the faint giants, there is a population of blue ($(G_\mathrm{BP}-G_\mathrm{RP})_0\lesssim 0.8$) giant stars that are significantly more metal-rich in Pristine. This can also be seen in the corresponding CMDs color-coded by $\FeH_\mathrm{Pristine}-\FeH_\mathrm{DESI}$ (upper panel of Fig.~\ref{fig:cmd2}). The metallicities of these stars are overestimated by Pristine, whose photometric metallicity model is optimized for the metal-poor (MP) regime. After we apply the metallicity cut of $[\FeH_{\rm Pristine} < -1$, the discrepancy disappears for the faint blue stars, as shown in the middle row of Fig.~\ref{fig:cmd2}. 

The histogram of metallicity differences in the MP regime are shown in the bottom panels of Fig.~\ref{fig:cmd2}. We measure only a small systematic offset of $\sim-0.15$ between the Pristine and DESI metallicities for DESI giants, with a dispersion of 0.22. The offset is $-0.21$ for the MP dwarfs, with a dispersion of 0.27. This offset is similar to the offset of $\sim0.2$ reported between DESI and the metal-poor SAGA sample by \citet{Koposov2026}. 

In Fig.~\ref{fig:color}, we plot the stars in common between DESI and Pristine that have large metallicity discrepancies ($\Delta\FeH_{\rm DESI - Pristine} >$1.0), color-coded by $\FeH_{\rm Pristine}$ for the giant and dwarf classification determined by DESI and for the bright and faint samples. Stars above the solar metallicity lines are highlighted with black circles, and those below the lines (\texttt{Flag\_training}/\texttt{Flag\_training\_dw}=1) are plotted with gray circles. There are significantly more stars in the faint sample ($G > 17.65$) that have large differences, which is not surprising as the metallicity uncertainties significantly increase as stars get fainter. 

The total number of dwarf stars in the cross-match according to DESI is about 20 times the number of giant stars, and there are more dwarf stars that have large metallicity discrepancies as seen in Fig.~\ref{fig:desi}. There are multiple sources from both surveys that could contribute to such discrepancies, \eg Pristine performs less well for metal-rich dwarfs (see Section~\ref{sec:desi}).

%%%%%%%%%%%%%%%%%%%%%%%%%%%%%%%%%%%%%%%%%%%%%%%%%%

\end{document}